\documentstyle[epsf]{mn}

\newif\ifAMStwofonts

\def\spose#1{\hbox to 0pt{#1\hss}}
\def\lta{\mathrel{\spose{\lower 3pt\hbox{$\mathchar"218$}}
     \raise 2.0pt\hbox{$\mathchar"13C$}}}
\def\gta{\mathrel{\spose{\lower 3pt\hbox{$\mathchar"218$}}
     \raise 2.0pt\hbox{$\mathchar"13E$}}}

\let\lesssim=\lta
\let\gtrsim=\gta

\def\msun{{\rm\,M_\odot}}

\def\kms{{\rm\,km\,s^{-1}}}
\def\pc{\,\hbox{\rm pc}}
\def\kpc{{\rm\,kpc}}

\mathchardef\star="313F

\def\gtrsim{\mathrel{\hbox{\rlap{\hbox{\lower4pt\hbox{$\sim$}}}\hbox{$>$}}}}
\def\mmatrix#1{\null\,\vcenter{\normalbaselines
    \ialign{\hfil$##$\hfil\,\,&&\hbox to1pt{\mathstrut\hss\vrule\hss}\,\,\hfil$##$\hfil\crcr
      \mathstrut\crcr\noalign{\kern-\baselineskip}
      #1\crcr\mathstrut\crcr\noalign{\kern-\baselineskip}}}\,}

\def\pmmatrix#1{\left(\mmatrix{#1}\right)}

\newdimen\colsize
\colsize=\textwidth
\divide\colsize by 2
\advance\colsize-12.1pt

\ifoldfss
  \ifCUPmtlplainloaded \else
    \NewTextAlphabet{textbfit} {cmbxti10} {}
    \NewTextAlphabet{textbfss} {cmssbx10} {}
    \NewMathAlphabet{mathbfit} {cmbxti10} {} 
    \NewMathAlphabet{mathbfss} {cmssbx10} {} 
  \fi
  \ifAMStwofonts
    \ifCUPmtlplainloaded \else
      \NewSymbolFont{upmath} {eurm10}
      \NewSymbolFont{AMSa} {msam10}
      \NewMathSymbol{\upi}     {0}{upmath}{19}
      \NewMathSymbol{\umu}     {0}{upmath}{16}
      \NewMathSymbol{\upartial}{0}{upmath}{40}
      \NewMathSymbol{\leqslant}{3}{AMSa}{36}
      \NewMathSymbol{\geqslant}{3}{AMSa}{3E}

       \let\le=\leqslant
       \let\ge=\geqslant
    \fi
  \fi
\fi 

\ifnfssone
  \newmathalphabet{\mathit}
  \addtoversion{normal}{\mathit}{cmr}{m}{it}
  \addtoversion{bold}{\mathit}{cmr}{bx}{it}
  \newmathalphabet{\mathbfit} 
  \addtoversion{normal}{\mathbfit}{cmr}{bx}{it}
  \addtoversion{bold}{\mathbfit}{cmr}{bx}{it}
  \newmathalphabet{\mathbfss} 
  \addtoversion{normal}{\mathbfss}{cmss}{bx}{n}
  \addtoversion{bold}{\mathbfss}{cmss}{bx}{n}
  \ifAMStwofonts
    \ifCUPmtlplainloaded \else
      \UseAMStwoboldmath
      \makeatletter
      \new@mathgroup\upmath@group
      \define@mathgroup\mv@normal\upmath@group{eur}{m}{n}
      \define@mathgroup\mv@bold\upmath@group{eur}{b}{n}
      \edef\UPM{\hexnumber\upmath@group}
      \new@mathgroup\amsa@group
      \define@mathgroup\mv@normal\amsa@group{msa}{m}{n}
      \define@mathgroup\mv@bold\amsa@group{msa}{m}{n}
      \edef\AMSa{\hexnumber\amsa@group}
      \makeatother
      \mathchardef\upi="0\UPM19
      \mathchardef\umu="0\UPM16
      \mathchardef\upartial="0\UPM40
      \mathchardef\leqslant="3\AMSa36
      \mathchardef\geqslant="3\AMSa3E

       \let\le=\leqslant
       \let\ge=\geqslant
    \fi
  \fi
\fi 

\ifnfsstwo
  \DeclareMathAlphabet{\mathbfit}{OT1}{cmr}{bx}{it}
  \SetMathAlphabet\mathbfit{bold}{OT1}{cmr}{bx}{it}
  \DeclareMathAlphabet{\mathbfss}{OT1}{cmss}{bx}{n}
  \SetMathAlphabet\mathbfss{bold}{OT1}{cmss}{bx}{n}
  \ifAMStwofonts
    \ifCUPmtlplainloaded \else
      \DeclareSymbolFont{UPM}{U}{eur}{m}{n}
      \SetSymbolFont{UPM}{bold}{U}{eur}{b}{n}
      \DeclareSymbolFont{AMSa}{U}{msa}{m}{n}
      \DeclareMathSymbol{\upi}{0}{UPM}{"19}
      \DeclareMathSymbol{\umu}{0}{UPM}{"16}
      \DeclareMathSymbol{\upartial}{0}{UPM}{"40}
      \DeclareMathSymbol{\leqslant}{3}{AMSa}{"36}
      \DeclareMathSymbol{\geqslant}{3}{AMSa}{"3E}

       \let\le=\leqslant
       \let\ge=\geqslant
    \fi
  \fi
\fi 

\ifCUPmtlplainloaded \else
  \ifAMStwofonts \else 
    \def\upi{\pi}
    \def\umu{\mu}
    \def\upartial{\partial}
  \fi
\fi

\title{Dynamics of an interacting luminous disk, dark halo, and
  satellite companion}

\author[Martin D. Weinberg]{Martin D. Weinberg\thanks{Alfred P. Sloan
  Foundation Fellow.\newline e-mail: weinberg@phast.umass.edu}\\
  Department of Physics and Astronomy, University of Massachusetts,
  Amherst, MA 01003-4525, USA}

\date{}

\pagerange{\pageref{firstpage}--\pageref{lastpage}}

\pubyear{1997}

\begin{document}

\maketitle

\label{firstpage}

\begin{abstract}
  
  This paper describes a method for determining the dynamical
  interaction between extended halo and spheroid components and an
  environmental disturbance. One finds that resonant interaction
  between a galaxy and passing interlopers or satellite companions can
  carry the disturbance inward, deep inside the halo, where it can
  perturb the disk.
  
  Applied to the Milky Way for example, the LMC and SMC appear to be
  sufficient to cause the observed Galactic warp and possibly seed
  other asymmetries.  This is a multi-scale interaction in which the
  halo wake has a feature at roughly half the satellite orbital radius
  due to a 2:1 orbital resonance.  The rotating disturbance then
  excites an $m=1$ vertical disk mode which has the classic
  integral-sign morphology. A polar satellite orbit produces the
  largest warp and therefore the inferred LMC orbit is nearly optimal
  for maximum warp production.
  
  Both the magnitude and morphology of the response depend on the
  details of the disk and halo models.  Most critically, a change in
  the halo profile will shift the resonant frequencies and response
  location and consequently alter the coupling to the bending disk.
  Increasing the halo support relative to the disk, a sub-maximal disk
  model, decreases the warp amplitude.
    
  Finally, the results and prognosis for N-body simulations are
  discussed.  Discreteness noise in the halo, similar to that due to a
  population of $10^6M_\odot$ black holes, can produce observable
  warping.  Details are discussed in an associated paper.

\end{abstract}
  
\begin{keywords}
  Galaxy: halo, structure---galaxies: halos, kinematics and
  dynamics---Magellanic Clouds
\end{keywords}

\section{Introduction} \label{sec:intro}

Even casual examination shows that most disk galaxies are not truly
symmetric but exhibit a variety of morphological peculiarities of
which spiral arms and bars are the most pronounced.  After decades of
effort, we know that these features may be driven by environmental
disturbance acting directly on the disk, in addition to
self-excitation of a local disturbance (e.g.\ by swing amplification,
Toomre 1981, Sellwood \& Carlberg 1984\nocite{Toom:81,SeCa:84}).
However, all disks are embedded within halos and therefore are not
dynamically independent and will respond to asymmetries and
distortions in the halo, as well.

Until recently, conventional wisdom was that halos acted to stabilize
disks but otherwise remained relatively inert.  The argument behind
this assumption is as follows.  Halos, spheroids and bulges are
supported against their own gravity by the random motion of their
stars---a so-called ``hot'' distribution (e.g. Binney \& Tremaine
1987\nocite{BiTr:87}).  On all but the largest scales, they look like
nearly homogeneous thermal baths of stars.  Because all
self-sustaining patterns or waves in a {\it homogeneous}\/ universe of
stars with a Maxwellian velocity distribution are predicted to damp
quickly (e.g. Ikeuchi, Nakamura \& Takahara 1974\nocite{IkNT:74}), one
expects that any pattern will be strongly damped in halos and
spheroids as well.  However, recent work suggests that halos {\it
  do}\/ respond to tidal encounters by companions or cluster members
and {\it are}\/ susceptible to induction of long-lived modes due to
their inhomogeneity.  These modes are at the largest scales for which
self gravity is most effective.  In particular, if halos are large as
many currently estimate, halo-halo interactions in groups or clusters
will be frequent and much more common than disk-disk interactions.
Because non-local coupling can carry a disturbance to small radii,
transient halo structure may be sufficient to trigger disk structure
even if the halo pattern subsequently damps away.  The non-locality of
the response was predicted by Weinberg (1986,
1989\nocite{Wein:86,Wein:89}) and verified in n-body simulations by
Hernquist \& Weinberg (1989\nocite{HeWe:89}), and Prugniel \& Combes
(1992\nocite{PrCo:92}).  Application to self-gravitating fluctuations
is described in an associated paper (Weinberg 1997\nocite{Wein:97}).

Similarly, we expect that a companion can continuously re-excite
structure.  The orbit of the companion decays by dynamical friction.
The response of the dark halo to the interloping satellite can be
thought of as a gravitational wake (e.g. Mulder 1983\nocite{Muld:83});
since the wake trails and has mass, it exerts a backward pull on the
satellite.  This view of dynamical friction reproduces the standard
approach but shows that the halo responds with structure whose mass
comparable to the satellite.  This general approach has been tested
and compared with n-body simulations in a variety of contexts with
good agreement.  Recent work (Weinberg 1995b, Paper
I\nocite{Wein:95b}) suggests that the Magellanic Clouds use this
mechanism to produce distortions in the Galactic disk sufficient to
account for both the radial location, position angle and sign of the
HI warp and observed anomalies in stellar kinematics.  In clusters,
this mechanism is mostly likely the culprit behind {\it galaxy
  harassment} (Moore et al. 1996\nocite{MKLD:96}).

Here, we develop this suggestion and present a formalism for exploring
the dynamics of low-amplitude interactions that can lead to
significant long-term evolution.  As an example throughout, I will
focus on disk warping and on the Milky Way---LMC interaction presented
in Paper I.  Although there are some general trends, a relatively
large amplitude disk response tends to be the result of a conspiracy
between frequencies.  Rather than present an exhaustive parameter
survey, we explore some simple scenarios.  Even though the warp
amplitudes vary with the details of the galaxy profiles,
astronomically interesting amplitudes are generally produced.
Sections \ref{sec:model} and \ref{sec:method} describe the galaxian
models and the method.  We will use a numerical perturbation theory
which is well suited to describing weak coherent perturbations.  The
main results are in \S\ref{sec:results} which begins by examining an
example distortion of the halo by a companion satellite and traces its
influence on both the disk warp and in-plane disk distortions.
Readers may skip the technical detail without loss of continuity by
turning to \S\ref{sec:results} after the introduction to
\S{sec:method}.  The results section is followed by a discussion of
the range of effects using other models and rough generalizations
(\S\ref{sec:disc}).  It is tempting and desirable to follow this up
with n-body simulation but simulations of weak distortions with large
dynamic range is a significant challenge and will require very large
particle numbers to recover signal (\S\ref{sec:n-body}).  We end with
a summary and outline for future work (\S\ref{sec:summary}).

\section{Galaxy models}\label{sec:model}

In order to explore interactions between components, we need
self-consistent multi-component galaxy models.  Fully self-consistent
disk-in-halo models are generally made prescriptively rather than
constructively because of the difficulty in determining distribution
functions.  For example, a regular disk profile embedded in an a
regular spherical halo will generally not yield a self-consistent
regular system demanding computationally intensive techniques such as
Schwarzschild's method (1979\nocite{Schw:79}).  Most often in the
literature, the Jeans' moment equations are used to construct an
n-body disk in a given halo or spheroid and the resulting distribution
is allowed to phase mix to equilibrium.  In this paper, we adopt a
hybrid approach suited to both the analytic perturbation theory
described in \S\ref{sec:method} and the n-body simulations described
in \S\ref{sec:n-body}.  The prescription is as follows:
\begin{enumerate}
\item Choose a spherical halo and two- or three-dimensional disk
  profile.
\item Assume that the disk does not affect the halo profile and
  construct a disk phase-space distribution function for the disk in
  the halo.  This would rigorously obtain for $M_{halo}\gg M_{disk}$.
  The disk distribution function is computed by a quadratic
  programming scheme similar to that discussed by Dejonghe
  (1989\nocite{Dejo:89}, see Appendix \ref{sec:diskDF}).  The term
  halo here includes the entire {\it hot}\/ component: bulge, stellar
  spheroid and dark matter.  The halo distribution function is assumed
  to be known.  If it is not, it may be constructed using any of the
  established techniques (e.g. Eddington inversion, generalized
  integral inversion, atlas method, etc.), ignoring the disk.
\item The approximate distribution functions for each component are
  now known.  For an n-body simulation, these may be directly realized
  by a Monte Carlo procedure.
\end{enumerate}
The realized phase space distribution will not be in strict
equilibrium.  However, as long as the force of the halo dominates the
disk at large radii, and the disk dominates its own gravity inside its
scale length for realistic parameters, the initial construction is
mostly likely close to equilibrium.  Numerical experiments support
this conjecture.  Nevertheless, a mild inconsistency is of minor
consequence for the numerical method presented in \S\ref{sec:method}
(see discussion in \S\ref{sec:combined}).

\begin{figure*}
\mbox{
\mbox{\epsfxsize=\colsize\epsfbox{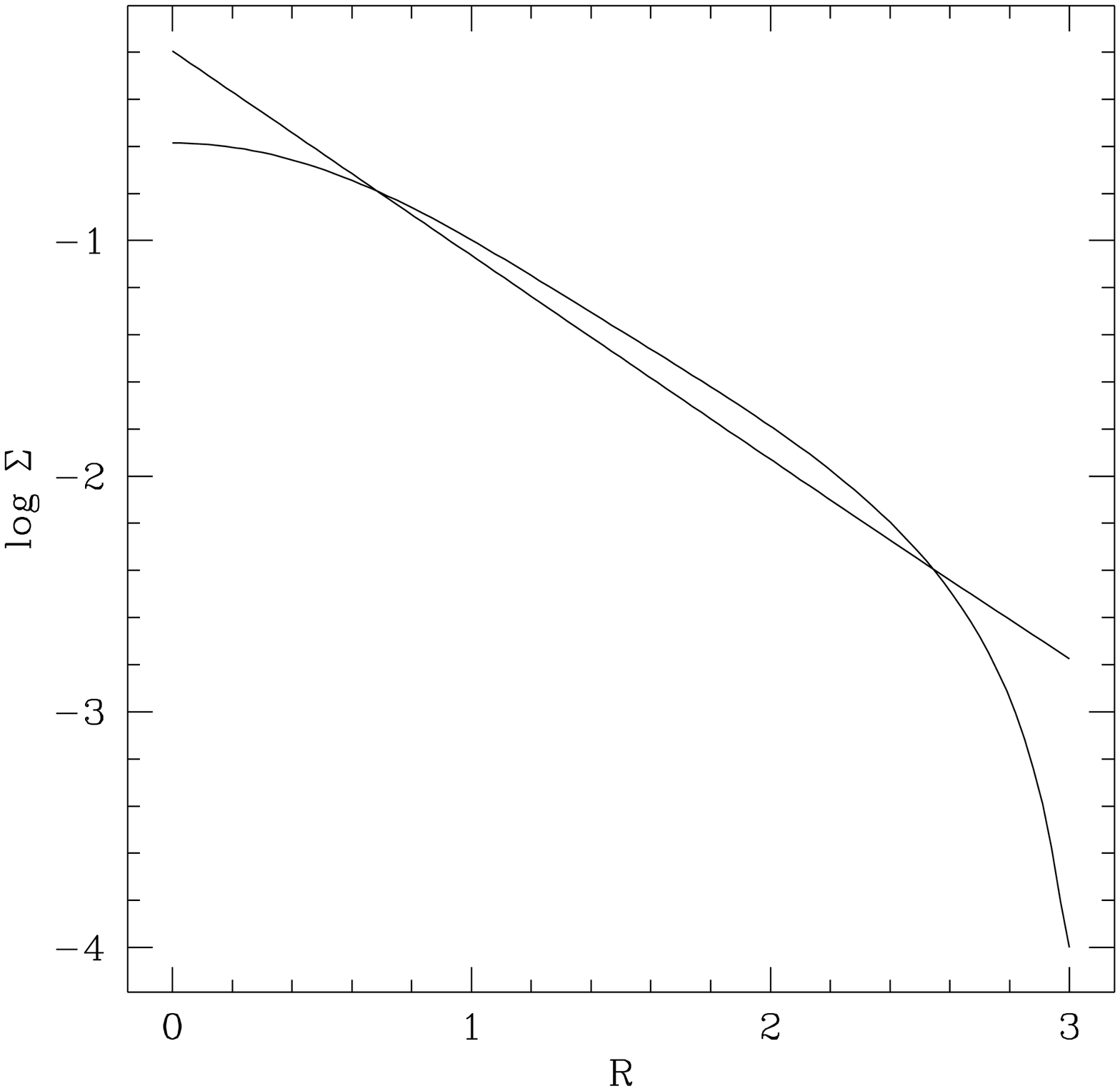}}
\mbox{\epsfxsize=\colsize\epsfbox{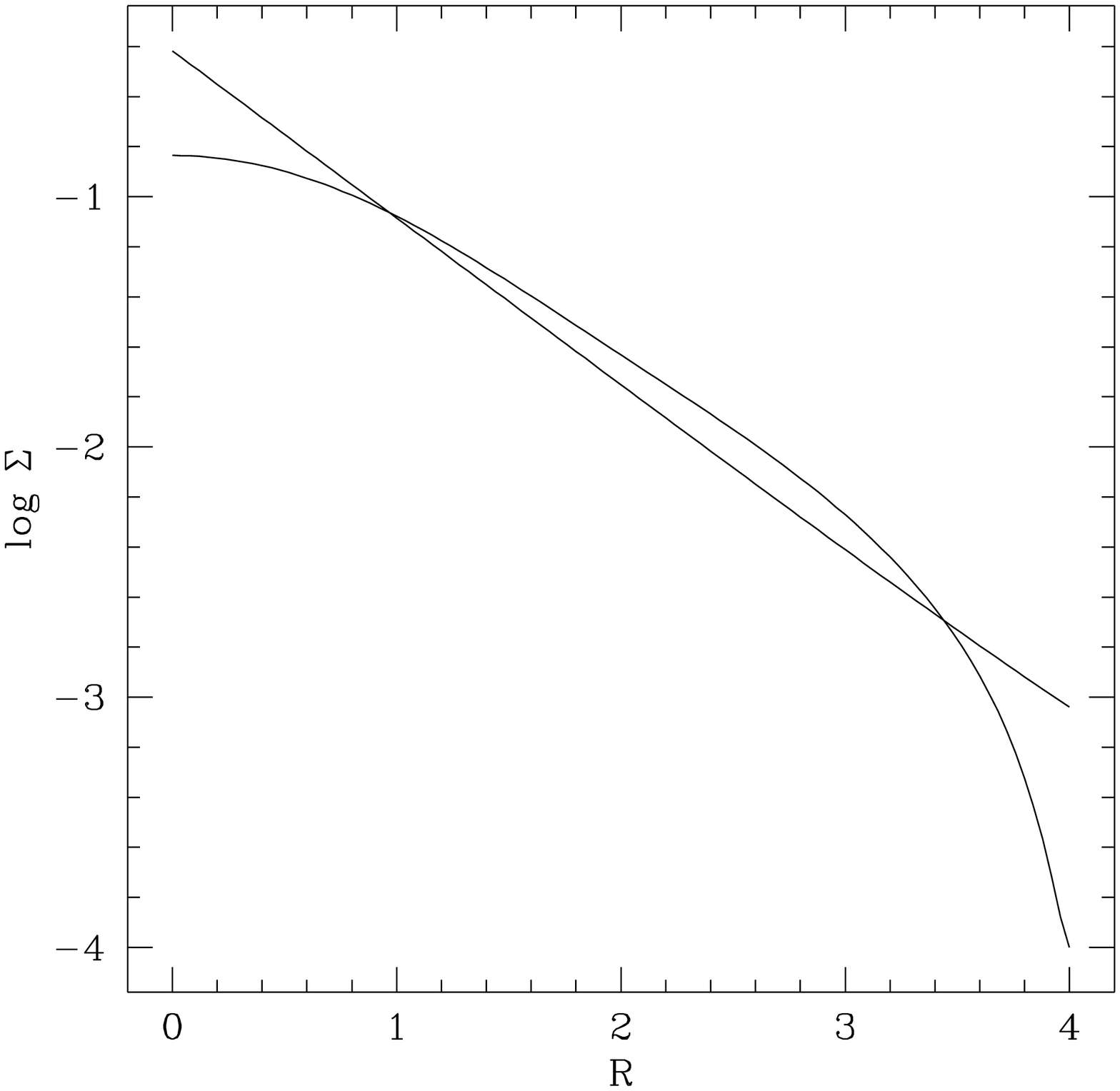}}
}
\caption{Left-hand (right-hand) panel compares the exponential disk to
  the Hunter-Toomre 16X model with identical mass inside of $R=3$
  ($R=4$) corresponding to approximately $21\kpc$ ($28\kpc$).}
\label{fig:exphX}
\end{figure*}

For analytic convenience, we use King models of varying concentration
and scale to represent the dark halo, a Hernquist model to represent
the bulge/spheroid, and Hunter (1963\nocite{Hunt:63}) polynomial disk
profiles.  The latter choice is motivated by the numerical analysis
required to compute the vertical disk response following Hunter \&
Toomre (1969\nocite{HuTo:69}, hereafter HT).  Hunter's polynomial disk
models are less centrally concentrated and fall of more steeply then
the standard exponential disk.  Hunter \& Toomre modified these disks
to better fit observed profiles by subtracting off a low-order
contribution; they denoted these models with the suffix ``X''.  The
Hunter-Toomre 16X model is fair approximation to the exponential disk
with a scale length of $R_{max}/6$; this corresponds to $a=3.5,
4.5\kpc$ for $R_{max}=21, 28\kpc$.  These profiles are compared in
Figure \ref{fig:exphX}.  For the n-body tests described in
\S\ref{sec:n-body}, we used a rigid bulge component to stabilize the
inner disk (which is otherwise bar unstable).

The extent of Milky Way disk remains under debate.  Robin, Cr\'ez\'e
\& Mohan (1993\nocite{RoCM:93}) present evidence for an edge to the
stellar disk at 14 kpc.  This is consistent with estimates based on
molecular tracers (Wouterloot et al. 1990, Digel 1991, Heyer et al.
1997\nocite{WBBK:90,Dige:91}) which imply an edge of about 14 kpc or
somewhat greater.  The outer atomic gas matches on to the inner
component continues to 30 kpc at significant surface densities
(Kulkarni, Heiles \& Blitz 1982\nocite{KuHB:82}).  A more recent
analysis (Diplas \& Savage 1991 \nocite{DiSa:91}) reports an
exponentially distributed neutral hydrogen disk out to {\it at least}
30 kpc.  We will adopt the $R_{max}=28\kpc$ model because it better
represents the global extent of the disk even though the scale length
required is a bit large.  In calculations below, the disk model is
assigned $M=1$ and this corresponds to the estimated
$6.0\times10^{10}\msun$ disk (Binney \& Tremaine
1987\nocite{BiTr:87}).  The radial scaling is one unit for each
$7\kpc$.  The satellite orbit is defined by its energy and angular
momentum in the halo model.  For each halo model, the orbit is
assigned a pericenter at $50\kpc$ and apocenter at $100\kpc$.  The
mass of the LMC in these units is 0.25 (0.1) for
$M_{LMC}=1.5\times10^{10}\msun\,(6\times10^9\msun)$ (Meatheringham
1988, Schommer et al. 1992\nocite{MDFW:88,SOSH:92}).  For this study,
we lump the Small Cloud together with the LMC and do not consider the
possibility of a distinct SMC orbit.  In all that follows, we will
scale results to the Milky Way.

Given the disk profile, the King radius and mass are constrained by
the demand for a flat rotation curve, at least for $R\lesssim50\kpc$.
The choice of a $W_0=3$ model with truncation radius at 200 kpc and
mass of 10 disk masses results in a Milky Way mass of
$3.3\times10^{11}$ inside of 50 kpc.  This is a bit smaller than but
comparable to current values; e.g. Kochanek (1996\nocite{Koch:96})
estimates $4.9\times10^{11}\msun$.  We also will consider halo mass
ratios of 15 and 20 disk masses.  The rotation curves remain
approximately flat for both cases and give masses within 50 kpc of
$4.7\times10^{11}\msun$ and $6.0\times10^{11}\msun$, respectively.
Finally, the timing argument and its recent generalizations (Peebles
1995\nocite{Peeb:95} and references therein) estimate a Milky Way mass
of $2\times10^{12}\msun$.  This is a {\it constraint}\/ not a target
for our models since one can add mass beyond 100 kpc without affecting
any of the dynamical arguments considered here.  Our most massive
model is below this limit with a mass of $1.3\times10^{12}\msun$.

The perturbing satellite orbit is chosen to match the LMC orbit
inferred by Lin et al. (1995\nocite{LiJK:95}) and assigned an orbit
consistent with the chosen halo profile and scaling given in the
previous section.  The clouds are assumed to be near pericenter with
$R_{LMC}=49.5$ kpc.  The space velocity inferred from proper motion,
radial velocity measurement, and a distance estimate immediately
determines the orbital plane as described in Paper I.  In a
Galactocentric coordinate system with the LSR along the $x$-axis
($x=-8.5\kpc$) and moving towards positive $y$, one may
straightforwardly derive the following instantaneous orbit:
$-76\pm13^\circ$ inclination, $-82\pm10^\circ$ longitude of ascending
node, $-36\pm3^\circ$ argument of perigalacticon.  However, considered
in these Galactic coordinates, the Milky Way disk rotates clockwise
and has negative angular momentum.  In the development below, disks
are assigned positive angular momentum with counter-clockwise
rotation.  Our models may be transformed to the Milky Way system by
reflection through the $y$--$z$ plane.

We will adopt a standard galaxy model with a King model halo with
tidal radius $R_t=28$ (200 kpc), $\log c=0.67$, and a mass of 10 times
the disk mass; this is ``maximal'' disk model.  The bulge is Hernquist
model with scale length 0.2 (1.4 kpc) and mass of 20\% of the disk
($1.2\times10^{10}\msun$). The disk is the Hunter-Toomre 16X model
with $R_{max}=4$ (28 kpc). In this system, one velocity unit is
$350\kms$.  The rotation curve for this model rises from 0.6 at
$R=0.5$ to 0.7 at $R=1.8$, drops slowly to 0.6 at $R=10$ (70 kpc) and
drops off more rapidly beyond this point.  Although better fits to the
observed Milky Way rotation curve are available, our goal of
understanding the underlying mechanism and the computational
simplicity of these components supports our choice.  Finally, the
standard model includes a satellite with an LMC orbit.  The magnitude
of the response scales with satellite mass and need not be chosen a
priori.

\section{A formalism for multi-scale interactions} \label{sec:method}

A full treatment requires the dynamical coupling of the multiple time
scales and multiple length scales of the external disturbance and the
galaxian components discussed above.  Relevant characteristic length
and time scales may differ by an order of magnitude between satellite
and halo or disk orbits.  In addition, we will see that the halo
disturbance may be relatively weak and a significant perturbation of
the outer disk at the same time.  These multiple-scale weak regimes
are a challenging task for an n-body computation.  However, this class
of problems is ideally suited to linear techniques and the work here
will use the expansion technique known as the {\it matrix method}.
Although the matrix method is computationally intensive, it is no more
so than n-body methods and is practical on current workstations.  In
this section, I will give a brief overview of the general method with
details on posing and implementing the coupled response solutions in
the references cited below and in the Appendix.

In short, the matrix method represents the response of a galaxy to an
external perturbation by a truncated series of orthogonal functions,
similar to those one would use to solve an electrostatics problem.
The perturbation is also represented by this series and the temporal
dependence of each coefficient is Fourier transformed to a (complex)
frequency distribution.  The response of the galaxy to one of the
orthogonal functions at a particular forcing frequency is then
computed in the continuum limit using the collisionless Boltzmann
(Vlasov) equation.  The entire procedure is analogous to signal
processing in Fourier space.  Pursuing the analogy, we now do the
inverse transform. The response to any perturbation, the weighted
superposition of the response to each basis function, is then a matrix
equation.  Finally, to get the full time dependence of the response,
one resums the solutions to the matrix equation at each frequency
weighted by the Fourier coefficients.

This method assumes that the perturbation is small enough that the
overall change to the structure of the galaxy is small.  In this
limit, the method has the advantage of accuracy and sensitivity to the
large scale structures of interest.  For contrast, the n-body
simulation determines the response of a galaxy to a perturbation by
solving the equations of motion for a representative set of orbits.
The orbit is advanced in a fixed potential for a short time interval
and the gravitational potential or force is then recomputed.  The
simulation works well for large perturbations but because the
simulation uses a finite number of particles, fluctuation noise limits
the sensitivity to small amplitude distortions.  The matrix method
nicely complements the n-body simulations, excelling in the regimes
where the n-body simulation are suspect.

Historically, the approach is related to the treatment of general
eigenvalue problems described in the mathematical physics literature
(e.g. Courant \& Hilbert 1953, Chap V\nocite{CoHi:53}).  The matrix
method in stellar dynamics had varied applications beginning with
Kalnajs (1977\nocite{Kaln:77}) who investigated the unstable modes of
stellar disks.  Polyachenko \& Shukhman (1981\nocite{PoSh:81}) adapted
the method to study a spherical system (see also Fridman \&
Polyachenko 1984, Appendix\nocite{FrPo:84b}) and it was later employed
by both Palmer \& Papaloizou (1987\nocite{PaPa:87}) in the study of
the radial orbit instability and by Bertin \& Pegoraro
(1989\nocite{BePe:89}) to study the instability of a family of models
proposed by Bertin \& Stiavelli (1984\nocite{BeSt:84}).  In addition
to Paper I, Weinberg (1989\nocite{Wein:89}, Paper II) used the matrix
formulation to study the response of a spherical galaxy to an
encounter with a dwarf companion and Saha (1991\nocite{Saha:91}) and
Weinberg (1991\nocite{Wein:91a}) investigated the stability of
anisotropic galaxian models.

\subsection{Mathematical overview} \label{sec:overview}

The response of a galaxy initially in equilibrium to a gravitational
interaction with a companion is described by the simultaneous solution
of the Boltzmann and Poisson equations.  The simultaneous system is a
set of coupled partial integro-differential equations.  However if the
orbits in each component are regular, any phase-space quantity---such
as density and gravitational potential---may be expanded in a Fourier
series in the orbital frequencies.  Truncating this expansion, the
quantity may be represented as a vector of Fourier coefficients; this
is standard practice in filtering and approximation theory and
canonical perturbation theory (e.g. Lichtenberg \& Lieberman
1983\nocite{LiLi:83}).  In Fourier space, the Boltzmann PDE becomes an
algebraic integral equation.  The system is further simplified if the
basis functions are chosen to satisfy the Poisson equation explicitly.
After a Laplace transform in time, the remaining solution of the
Boltzmann equation becomes the solution of a matrix equation, each
column describing the response to a particular basis function.  See
references cited abouve for mathematical detail.

To give the flavor of its use, consider two interacting galaxies.
Denote the expansion of the perturbation potential caused by the
companion galaxy as vector ${\bf b}$.  Then the direct response of the
galaxy, vector ${\bf a}$ may be written:
\begin{equation} \label{eq:resp}
        {\bf a} = {\bf R}\cdot{\bf b}.
\end{equation}
The matrix ${\bf R}$, the {\it response operator}, implicitly contains
the time-dependence of the perturbation as well as the dynamics of the
Boltzmann equation.  If we are interested in the self-gravitating
response to the perturbation, we need to solve:
\begin{equation} \label{eq:mat}
        {\bf a} = {\bf R}\cdot\left({\bf a} + {\bf b}\right).
\end{equation}
In words, equation (\ref{eq:mat}) states that the self-consistent
reaction of a galaxy to a perturbation is the gravitational response
to both the perturbing force and the force of the response itself.
Equations (\ref{eq:resp}) and (\ref{eq:mat}) result from the Laplace
transform of the Boltzmann equation and therefore represent a
particular frequency component, ${\bf a}={\bf a}(s)$.  Therefore, the
solution of equations (\ref{eq:resp}) or (\ref{eq:mat}) requires an
inverse Laplace transform to recover its explicit time dependence (see
Paper II for details).  See Nelson \& Tremaine (1997\nocite{NeTr:97})
for a general discussion of the response operator.

\subsection{Combining multiple components} \label{sec:combined}

This approach is easily extended to find the simultaneous
self-consistent response of several galaxian components.  For example,
let us consider a halo and a disk; any number of components may be
combined similarly.  Each component's distribution function solves a
Boltzmann equation and is coupled to the others only through the total
gravitational potential.  If the interaction between the components is
artificially suppressed, the simultaneous solution is that of the
following augmented matrix equation:
\begin{equation} \label{eq:augmat}
\pmatrix{{\bf a}_h\cr{\bf a}_d} =
        \pmmatrix{
        {\bf R}_h & {\bf 0}\cr\noalign{\hrule}
        {\bf 0} & {\bf R}_d\cr} \cdot
        \left[
        \pmatrix{{\bf a}_h\cr{\bf a}_d} +
        \pmatrix{{\bf b}_h\cr{\bf b}_d} \right],
\end{equation}
where the subscripts $h$ and $d$ stand for the halo and disk, {\bf b}
is the external perturbation, and ${\bf 0}$ is defined to be the
matrix with the same rank as ${\bf R}$ and all elements zero.
Equation (\ref{eq:augmat}) is simply two stacked versions of equation
(\ref{eq:mat}).  The off-diagonal partitions in the augmented matrix,
now ${\bf 0}$, describe the mutual interaction between components.

To allow the components to interact, we project the halo response
${\bf a}_h$ onto the disk basis and then add this to the
right-hand-side of the disk response equation and vice versa for the
response of the halo to the disk.  Letting the matrices that perform
these projections be ${\bf P}_{hd}$ and ${\bf P}_{dh}$, we may write
the fully coupled set as
\begin{eqnarray} \label{eq:coupled_matrix}
\pmatrix{{\bf a}_h\cr{\bf a}_d} &=&
        \pmmatrix{
        {\bf R}_h & {\bf 0}\cr\noalign{\hrule}
        {\bf 0} & {\bf R}_d\cr} \cdot
        \pmmatrix{
        {\bf 1} & {\bf P}_{hd}\cr\noalign{\hrule}
        {\bf P}_{dh} & {\bf 1}\cr} \cdot
        \pmatrix{{\bf a}_h\cr{\bf a}_d} +
        \nonumber \\ &&
        \pmmatrix{
        {\bf R}_h & {\bf 0}\cr\noalign{\hrule}
        {\bf 0} & {\bf R}_d\cr} \cdot
        \pmatrix{{\bf b}_h\cr{\bf b}_d},
                                        \nonumber\\ \noalign{\vspace{5pt}}
&=&
        \pmmatrix{
        {\bf R}_h & {\bf R}_h{\bf P}_{hd} \cr\noalign{\hrule}
        {\bf R}_d{\bf P}_{dh}& {\bf R}_d\cr} \cdot
        \pmatrix{{\bf a}_h\cr {\bf a}_d} +
        \nonumber \\ &&
        \pmmatrix{
        {\bf R}_h & {\bf 0}\cr\noalign{\hrule}
        {\bf 0} & {\bf R}_d\cr} \cdot
        \pmatrix{{\bf b}_h\cr{\bf b}_d}.
\end{eqnarray}
The first term on the right-hand-side describes the mutual
self-gravitating response due to each component and the second term
describes the response of each component to the external perturbation.

We may straightforwardly isolate effects of interest by coupling or
uncoupling components or by including or suppressing self-gravity.
For example, we may consider the response of the disk to a halo wake,
but without the back reaction of the halo to the disk by setting the
upper right term in the augmented response matrix to zero.  Or, if we
want to limit the disk response to forcing by the halo alone, the
direct response of the disk to the perturbation may be left out by
setting ${\bf b}_d$ to zero, yielding
\begin{equation}
\pmatrix{{\bf a}_h\cr{\bf a}_d} =
        \pmmatrix{
        {\bf R}_h & {\bf 0} \cr\noalign{\hrule}
        {\bf R}_d{\bf P}_{dh}& {\bf R}_d\cr} \cdot
        \pmatrix{{\bf a}_h\cr{\bf a}_d} +
        \pmmatrix{
        {\bf R}_h & {\bf 0}\cr\noalign{\hrule}
        {\bf 0} & {\bf R}_d\cr} \cdot
        \pmatrix{{\bf b}_h\cr{\bf 0}}.
\end{equation}
This is equivalent to first solving ${\bf a}_h = {\bf
  R}_h\cdot\left({\bf a}_h + {\bf b}_h\right)$ and then ${\bf a}_d =
{\bf R}_d\cdot\left({\bf a}_d + {\bf P}_{dh}{\bf a}_h\right)$.  This
formalism couples the components by their perturbation from
equilibrium, not the fully gravitational attraction.  For this reason,
any mild deviation from perfect self consistency produced by the
prescription described in \S\ref{sec:model} does not cause any
problems and is unlikely to be significant.  Most of the computational
work is in producing the matrices ${\bf R}$.  Afterward, all of the
variants may be studied with little additional effort.

\subsection{Satellite perturbation}

In order to apply the technique described in \S\ref{sec:combined} to a
perturbation by an orbiting satellite, we need to expand its potential
in the chosen basis to get the perturbation vector ${\bf b}$.  This
expansion is described in \S\ref{sec:expan} and applied in
\S\ref{sec:appresp}.  Perturbation by an interloping galaxy may be
treated similarly but is not described here.

\subsubsection{Fourier expansion}
\label{sec:expan}

Our complete set of basis functions are pairs of functions, $(p_i,
d_i)$, which solve Poisson's equation, $\nabla^2 d_i = 4\pi Gp_i$, and
are biorthogonal:
\begin{equation}
{1\over4\pi G}  \int d{\bf r}\, p^\ast_i({\bf r}) d_j({\bf r}) =
\delta_{ij}.
\end{equation}
The  potential for the arbitrary point mass may be expanded
directly in a biorthogonal harmonic series:
\begin{equation}\label{eq:satexp}
  \Phi({\bf r}) = \sum_{lm}Y_{lm}(\theta, \phi) \sum_i b_k^{lm}(t) p_i^{lm}(r)
\end{equation}
where 
\begin{eqnarray}
  b_i^{lm}(t) &=& \int d^3r Y^\ast_{lm}(\theta, \phi) p^{lm\,\ast}_i(r)
  \delta^3\left({\bf r} - {\bf r}(t)\right)  \nonumber \\ &=&
  Y^\ast_{lm}(\theta(t), \phi(t)) p^{lm\,\ast}_i(r(t))
  \label{eq:ptexpand}
\end{eqnarray}
where ${\bf r(t)}$ describes the orbit of the satellite.
Alternatively, since $b^{lm}_i$ is an implicit function of time
through ${\bf r}$, we may expand in an action-angle series which makes
the time dependence explicit.  This gives
\begin{equation} \label{eq:satfreq}
  b_i^{lm}(t) = \sum_{l_1, l_2=-\infty}^\infty e^{im\gamma}
  e^{il_2\alpha}
  Y_{lm}(\pi/2, 0) W^{l_1\,i\,\ast}_{l\,l_2\,m}
  e^{i(l_1\Omega_1+l_2\Omega_2)t}
\end{equation}
where the coefficient $W^{l_1\,i\,\ast}_{l\,l_2\,m}$ depends only on
the energy and angular momentum of the satellite orbit.  Derivation of
equation (\ref{eq:satfreq}) is given in the Appendix.

The perturbation vectors $b$ in equation (\ref{eq:mat}) are given by
the Laplace transform of $b_i^{lm}(t)$ from equation (\ref{eq:satexp})
for each term in equation (\ref{eq:satfreq}).  The Laplace transform
of $b_i^{lm}(t)$ in this form is trivial.  Because the physical
measurable must be real, we can simplify the analytic computation for
$m\not=0$ by using only $m>0$ terms and adding the complex conjugate
in the end.

\subsubsection{Application to response calculation} \label{sec:appresp}

Finally, recovery of the response, ${\bf a}$, in the time domain
requires the inverse Laplace transform of the following solution for
${\bf a}$ from equation (\ref{eq:coupled_matrix}):
\begin{eqnarray}
\pmatrix{{\bf a}_h\cr{\bf a}_d} &=&
        \left[
        \pmmatrix{
        {\bf 1} & {\bf 0}\cr\noalign{\hrule}
        {\bf 0} & {\bf 1}\cr} -
        \pmmatrix{
        {\bf R}_h & {\bf R}_h{\bf P}_{hd} \cr\noalign{\hrule}
        {\bf R}_d{\bf P}_{dh}& {\bf R}_d\cr}
        \right]^{-1} \cdot
        \nonumber \\ &&
        \pmmatrix{
        {\bf R}_h & {\bf 0}\cr\noalign{\hrule}
        {\bf 0} & {\bf R}_d\cr} \cdot
        \pmatrix{{\bf b}_h\cr{\bf b}_d} \nonumber \\ \noalign{\leftline{or, in
        compact form,}}
{\bf{\hat a}}^{lm}(s) &\equiv& {\bf\cal D}^{-1}(s) \cdot {\bf\cal R}(s)\cdot {\bf{\hat b}}^{lm}(s).
        \label{eq:matresp}
\end{eqnarray}
The matrix in large square brackets in this equation is the dispersion
relation; its determinant vanishes at eigenmodes.  The assumption that
our multi-component galaxy is stable ensures that the inverse of this
matrix has no poles in the complex half plane with $\Re(s)>0$; poles
on the half plane with $\Re(s)<0$ correspond to damped modes.
Elements of the second term matrix have at least one pole on the
imaginary $s$-axis due to the harmonic forcing by the satellite. In
the end, the inverse transform may be simply evaluated by deforming
the integration contour through the imaginary axis and taking the
time-asymptotic limit (see \S\ref{sec:numresp} for details).  After
many satellite orbits, the harmonic-forcing contribution dominates all
but a very weakly damped mode.

\subsection{Disk bending}

A differential vertical force applied by the external perturbation and
the halo wake can warp the disk plane.  HT describe a linearized
solution for the dynamical evolution of the plane for an isolated
disk.  The bending analysis here uses the formalism developed by HT
with several modifications.  First, because our equilibrium disk
models are embedded in an external halo, we must retain their equation
(12) rather than simplify using relationships based on the specific
form of the background model.  Second, we solve the linearized
equations of motion under a forced disturbance (their eq. 19) by
Laplace transform for consistency with the approach in
\S\ref{sec:combined}.  The vertical force follows directly from the
expansion coefficients describing the external perturbation, equation
(\ref{eq:satfreq}) and the halo wake, equation (\ref{eq:matresp}).
The back-reaction to the in-plane distortion is included and is a
relatively minor contribution to the total response.  This calculation
does not consider the back-reaction of the halo to the vertical
distortion (Nelson \& Tremaine 1995\nocite{NeTr:95}).

For reasons described in HT, their polynomial disk models are well
suited to numerical analysis and adopted here as described in
\S\ref{sec:model}.  I also tried exponential disks with different
basis sets but could not find an alternative which allowed an accurate
computation of the height alone, rather than the combining height
times the surface density.

\section{Wakes and warps} \label{sec:results}

\subsection{Wake in halo}

\begin{figure}
\mbox{\epsfxsize=\colsize\epsfbox{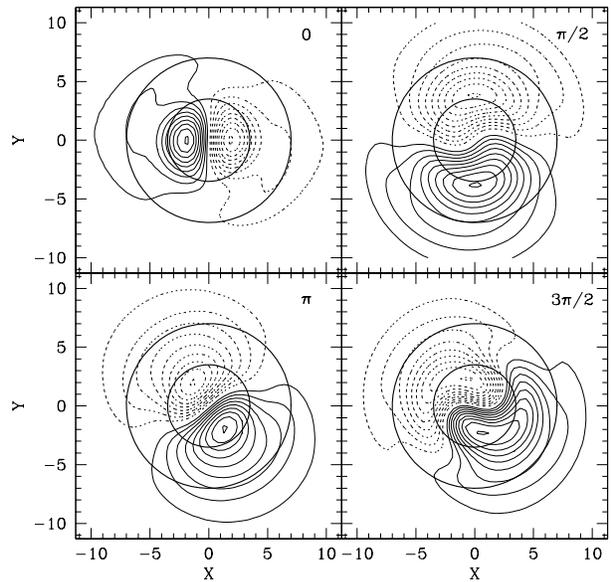}}
\caption{Wake in the fiducial halo due to orbiting satellite for
  $l=m=1$ (dipole).  Nine linear spaced contours of overdensity
  (solid) and underdensity (dotted) shown.  Each successive panel
  shows the wake radial phase as labeled, with pericenter at phase 0
  and apocenter at phase $\pi$.  Satellite locations at the four
  phases are $(X,Y) = (7.0, 0.0), (0.2, 11.6), (-8.5, 11.2), (-11.3,
  2.9)$.  The outer and inner circles indicate the pericentric and
  half-pericentric radii, respectively.}
\label{fig:halowake1}
\end{figure}

\begin{figure}
\mbox{\epsfxsize=\colsize\epsfbox{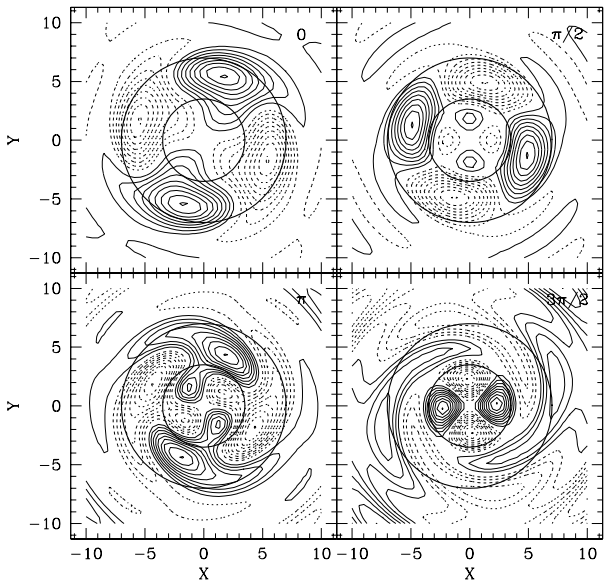}}
\caption{As in Fig. \protect{\ref{fig:halowake1}} but for $l=m=2$
  (quadrupole)halo}
\label{fig:halowake2}
\end{figure}

There are two contributions to the disk disturbance: the direct tidal
force of the satellite on the disk and the force of disturbance
excited by the satellite in the halo---the halo wake---on the disk.
The strength of the halo wake is proportional to the mass of the
satellite and mass density of the halo.  However it depends critically
on the orbital structure of the halo because a particle's orbit will
respond most strongly near resonances between the satellite's and halo
particle's orbital frequencies.  Co-orbiting trajectories will have
the strongest response but the total mass involved is small for the
standard model.  Higher-order resonances will be weaker but occur at
smaller galactocentric radii where the mass density is high.  The wake
is the product of these competing effects and, generally, the wake
peaks far inside the satellite orbit.

As an example, Figures~\ref{fig:halowake1} and \ref{fig:halowake2}
show the space density distortions induced by the LMC orbit in the
standard King model halo in the orbital plane.  The satellite has
pericenter at $R=7$ and apocenter at $R=14$ and here, orbits in the
counter-clockwise direction.  Pericenter is $X=7, Y=0$.  If the
satellite were completely outside the halo, the dipole response would
be a linear displacement representing the new center of mass position.
The wake would be proportional to $-d\rho(r)/d{\bf r}\cdot{\bf e}$
where ${\bf e}$ is the unit vector from the halo to the satellite
center (Weinberg 1989\nocite{Wein:89}).  In Figure
\ref{fig:halowake1}, the satellite is inside the halo, and the wake
deviates from the pure displacement.  The amplitude of the quadrupole
(Fig.  \ref{fig:halowake2}) is a factor of roughly two smaller than
the dipole.  The dominant wake is near the satellite pericenter as
expected, but note the inner lobe of the wake at roughly half the
pericenter distance.  This is due primarily to the 2:1 resonance
between satellite and halo orbital azimuthal frequencies.  Although
the relative density of this inner lobe is smaller than the primary
outer one at pericenter (phase 0), both the proximity and spatial
structure causes the force from the inner wake to dominate over direct
force from the satellite.  As the satellite approaches pericenter
(phase $3\pi/2$), these inner lobes become relatively stronger and can
dominate the response.  The inner galaxy wake is weaker past
pericenter (phase $\pi/2$).

\subsection{Vertical force on disk due to wake}

\begin{figure}
\mbox{\epsfxsize=\colsize\epsfbox{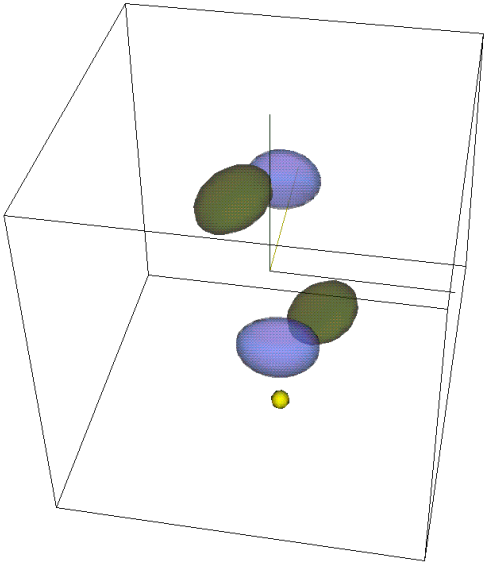}}
\caption{A three dimensional rendering of the $l=|m|=2$ 
  halo wake shown in Fig. \protect{\ref{fig:halowake2}}).  The
  isolevel shown at 75\% of the peak amplitude.  Overdensity and
  underdensity are shaded light grey and dark grey, respectively.  The
  wire-frame outline extends $\pm10$ units in $x$, $y$, and $z$ and
  the $x$--$y$--$z$ axes are shown with the $z$-direction along the
  vertical.  The satellite pericenter is a bit over $7$ units and
  shown as a small sphere.}
\label{fig:render}
\end{figure}

\begin{figure}
  \mbox{\epsfxsize=\colsize\epsfbox{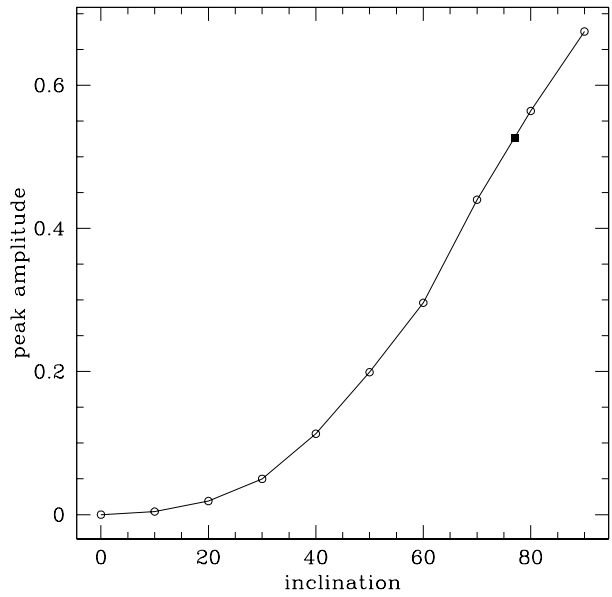}}
  \caption{Warp amplitude as a function of orbital plane
    inclination.  A polar orbit produces the maximum warp with
    heights 473 pc (1181 pc) for LMC mass of $6\times10^{9}\msun$
    ($1.5\times10^{10}\msun$).  The filled square shows the
    elevation of inferred LMC orbit.}
  \label{fig:elev}
\end{figure}

In the absence of the halo, the first multipole contributing to the
differential or {\it tidal} acceleration of the disk is at quadrupole
($l=2$) order.  It is straightforward to convince oneself of this
fact: the $l=0$ term is constant yielding no force, the $l=1$ term is
proportional to $r$ yielding a spatially constant force, and therefore
the $l=2$ term provides the lowest order differential force.  Because
the warp has $m=1$ symmetry, the dominant warp inducing term will be
$l=2$, $|m|=1$.  Similar symmetries apply for the action of the
perturbed halo on the disk.  Including the halo warp, the dipole still
can not produce the classic odd integral-sign warp but causes an even
distortion.  The lowest order halo wake that can excite an
integral-sign warp is also the $l=2, |m|=1$ component.  To illustrate
the three-dimensional structure, Figure \ref{fig:render} renders the
isosurface corresponding to 75\% of peak amplitude wake contoured in
Figure \ref{fig:halowake2}.  The wake is symmetric about the
satellite's orbital plane and this plane is easily visualized.  A wake
must asymmetric about the $z$ axis to cause a differential vertical
acceleration of the disk; in other words, a satellite in the disk
plane produces no warp.  The maximum vertical force occurs when the
pattern shown in Figures \ref{fig:halowake2} and \ref{fig:render} is
oriented perpendicular to the disk plane (see Fig. \ref{fig:elev}).
The orientation of the LMC disk plane is nearly ideal for producing a
disk warp.

\begin{figure}
\mbox{\epsfxsize=\colsize\epsfbox{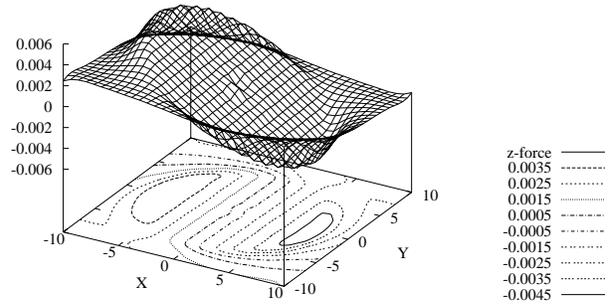}}
\caption{Vertical force on disk plane due to orbiting satellite only
  for harmonics $l=2, |m|=1$.  Shown midway between pericenter and
  apocenter, at radial phase $\pi/2$ (cf. Fig.
  \protect{\ref{fig:halowake1}}).}
\label{fig:vertforceL}
\end{figure}

\begin{figure}
\mbox{\epsfxsize=\colsize\epsfbox{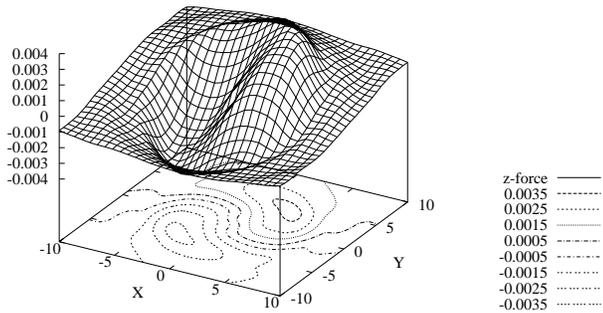}}
\caption{As in Fig. \protect{\ref{fig:vertforceL}} but 
  for the halo response to the satellite.
  }
\label{fig:vertforceR}
\end{figure}

Figures \ref{fig:vertforceL}--\ref{fig:vertforceR} describes the
vertical force on the disk plane due to both the satellite alone and
the halo response to the satellite for the standard model.  Although
the net force from the satellite is similar in magnitude to that from
the halo, the satellite force is varies linearly with distance and
only gives rise to a uniforming tilting of the disk.  However, the
spatial structure in the halo force causes a differential bending of
the disk with vertical $m=1$ symmetry: a warp.

\subsection{Vertical response of the disk}

\begin{figure}
  \mbox{\epsfxsize=\colsize\epsfbox{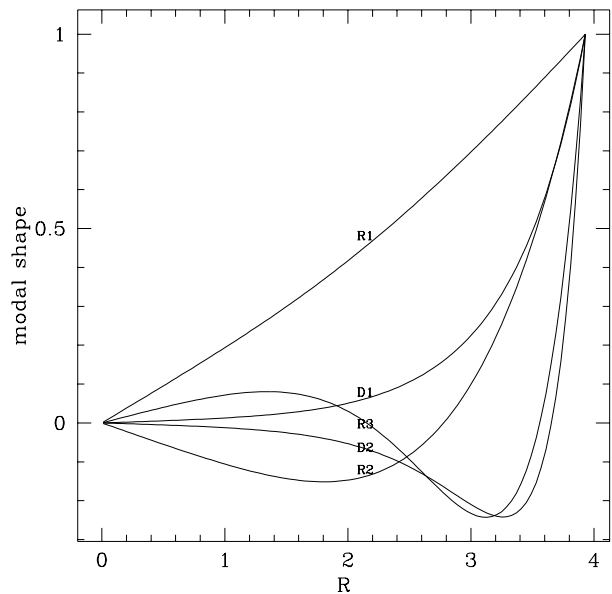}}
  \caption{Tipping modes for the Hunter $N=16$ disk with edge
    at $R=4$ (approximately $28\kpc$) embedded in a $W_0=3$ King model
    with $R_t=28$ (approximately $200\kpc$ scaled to the Milky Way).
    Modes are labeled in the notation of HT.}
  \label{fig:vmodes}
\end{figure}

\begin{figure}
  \mbox{\epsfxsize=\colsize\epsfbox{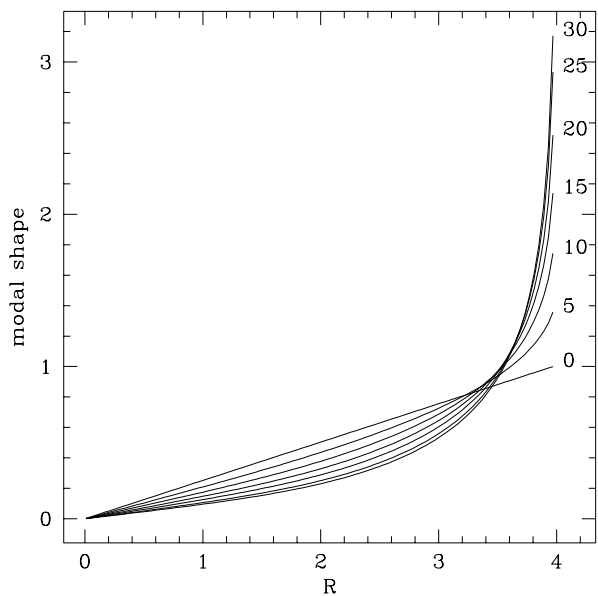}}
  \caption{Evolution of the tipping modes for the Hunter-Toomre 16X disk
    embedded in a $W_0=3$ King model with $R_t=28$ (approximately
    $200\kpc$ scaled to the Milky Way) as a function of halo mass.
    Curves are labeled by halo-to-disk mass ratio.  At zero halo mass,
    the mode is bodily tipping of the disk at zero frequency.  Each
    curve parallel to the $y$ axis describes the mode with an
    increasing halo mass fraction shown on the $x$ axis, up to a
    possible total of $M=30$ times the disk mass.}
\label{fig:modeevol}
\end{figure}

One can think of the vertical response as a superposition or packet of
modes.  Bending modes in the absence of a halo have been described by
HT.  The addition of the halo shifts and slightly modifies the shape
of these modes although they are qualitatively similar to those
described in Figure 3 of HT.  The corresponding modes in the standard
halo model ($W_0=3$ King model halo with truncation radii at $R_t=28$
[$200\kpc$]) are shown in Figure \ref{fig:vmodes}.  In the absence of
the halo, the zero frequency $m=1$ mode is a bodily tipping of the
disk.  As the halo mass increase relative to the disk mass, the shape
of the modes change.  The tipping mode, for example, evolves into a
distortion with warp-like structure.  For the standard (Hunter-Toomre
16X) disk, the tipping mode is the only discrete mode and appears to
be a prominent feature of the response discussed below.  Its shape for
increasing halo to disk mass ratio is shown in Figure
\ref{fig:modeevol}.  The tipping mode is distorted from linear to an
integral-sign shape.

\begin{figure}
  \mbox{\epsfxsize=\colsize\epsfbox{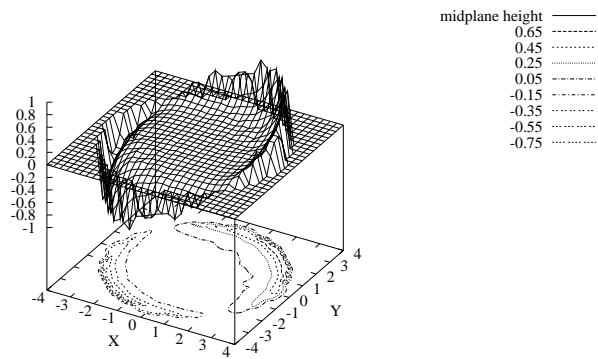}}
  \caption{Vertical response of the Hunter-Toomre 16X disk due to the
    combined vertical force from the standard satellite and halo.
    Scaled to the LMC and Milky Way following
    \S\protect{\ref{sec:model}}, the peak height is 420 pc (1050 pc)
    for the lower and higher LMC mass estimates, respectively.}
\label{fig:h432}
\end{figure}

The combination of the force exerted by the satellite and the
satellite-induced halo wake excites a vertical response in the disk.
A strong vertical disk response obtains for near commensurable
halo-wake pattern speeds and disk bending mode frequencies.  These are
{\it accidental} resonances in the sense that their existence is
circumstantial and not the result of a tuning mechanism.
Figure~\ref{fig:h432} shows the response to the combined force from
Figures \ref{fig:vertforceL}--\ref{fig:vertforceR}.  The vertical
distortion of the midplane within a galactocentric radius of roughly
$10\kpc$ is less than $100\pc$, however, the warp reaches a peak
height of about $400\pc$ ($1.0\kpc$) at $R_g\approx20\kpc$ for LMC
mass of $6\times10^9\msun$ ($1.5\times10^{10}\msun$).  In many cases,
the warp has a local maximum in the outer disk, corresponding to the
location maximum curvature in the integral sign.  At larger radii the
mode may turn over and reach a global maximum amplitude at the very
edge of the disk, corresponding to the ends points of the integral
sign.  We will refer to this first local maximum as the {\it peak}
when describing warps. The sharp edge has been truncated in Figure
\ref{fig:h432} and others below to best show this first maximum. We
will see below, that depending on the halo structure, both larger and
smaller warp distortions are possible with the same satellite
perturbation.

\subsection{In-plane response of the disk}

\begin{figure}
\mbox{\epsfxsize=\colsize\epsfbox{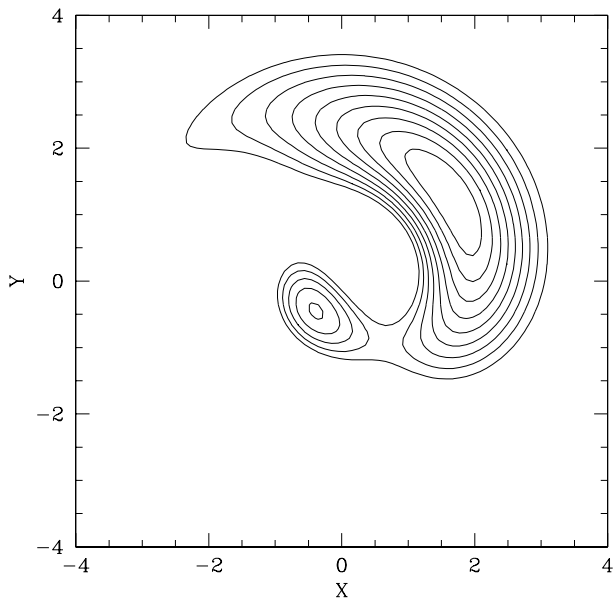}}
\caption{$m=1$ density distortion in the disk plane due to the
  satellite and halo perturbation.  Contours are evenly spaced from
  10\% to 90\% of maximum.}
\label{fig:inplane432m1L}
\end{figure}

\begin{figure}
\mbox{\epsfxsize=\colsize\epsfbox{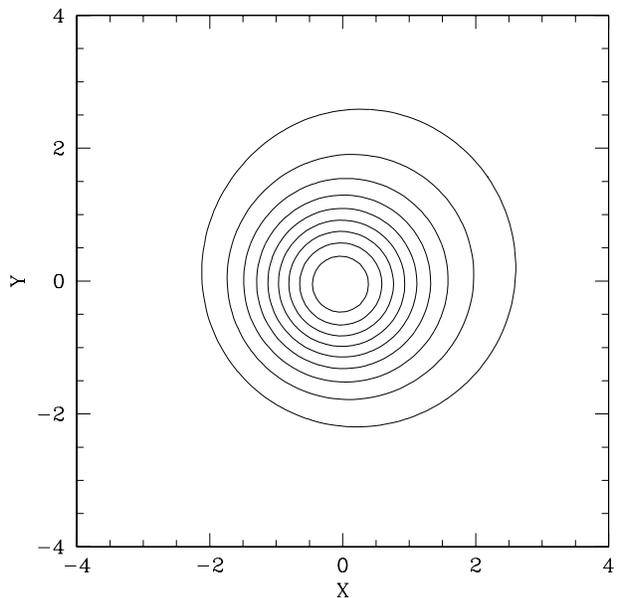}}
\caption{As in Fig. \protect{\ref{fig:inplane432m1L}} but 
  combined with the background for the larger LMC mass estimate
  (right).}
\label{fig:inplane432m1R}
\end{figure}

\begin{figure}
\mbox{\epsfxsize=\colsize\epsfbox{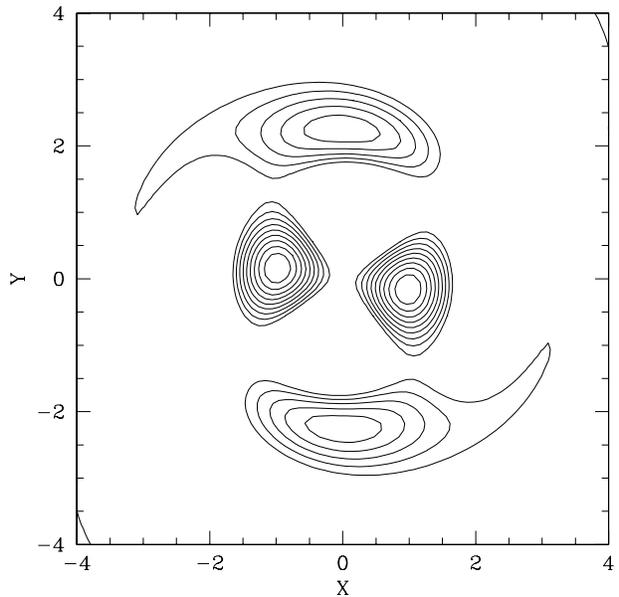}}
\caption{As in Fig. \protect{\ref{fig:inplane432m1R}} by for the $m=2$
  density distortion.}
\label{fig:inplane432m2L}
\end{figure}

\begin{figure}
\mbox{\epsfxsize=\colsize\epsfbox{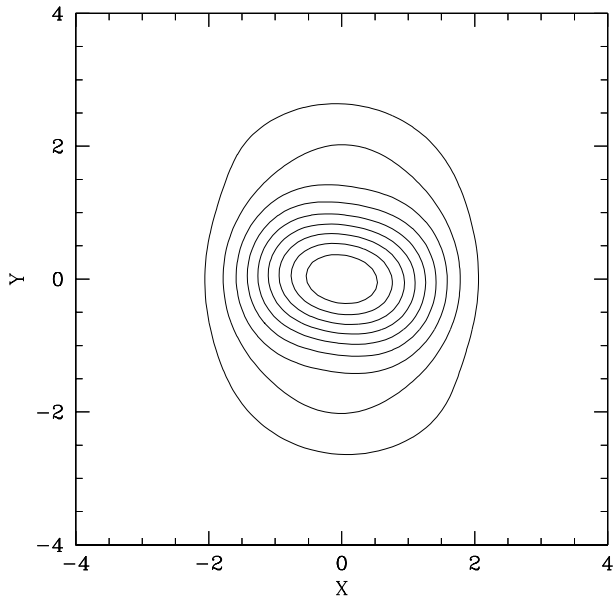}}
\caption{As in Fig. \protect{\ref{fig:inplane432m1L}} by for the $m=2$
  density distortion.  The mass of the LMC is exaggerated by a factor
  of 16 to illustrate the shape of the feature.}
\label{fig:inplane432m2R}
\end{figure}

To complete the example, we describe the concurrent in-plane
distortion for $m=1$ and $m=2$ harmonics.  The overall distortion is
small in the inner disk, $R_g\lesssim R_o=8.5\kpc$, and dominated by
the $m=1$ term with a relative density amplitude of roughly 1.6\%
(4\%) for the small (large) LMC mass estimate.  The $m=1$ distortion
appreciable in the outer disk, reaching 16\% (40\%) near
$R_g\gtrsim16\kpc$ and increasing beyond that.  This distortion will
produce an observably lopsided disk outer disk (see Figs.
\ref{fig:inplane432m1L}--\ref{fig:inplane432m1R}).  Unfortunately, it
is difficult to determine precise distances to gas in the outer galaxy
and this signature will be difficult to affirm.

The quadrupole leads to a measurable oval distortion only for
$R_g\gtrsim20\kpc$ of 6\% (16\%) and is at the percent level or
smaller near the solar circle (see Figs.
\ref{fig:inplane432m2L}--\ref{fig:inplane432m2R}).  This mild oval
distortion is much smaller than and will be swamped by the predicted
$m=1$ signature.

\section{Discussion} \label{sec:disc}

There are no simple formulae describing the warp amplitudes in general
because of the complexity of the interaction.  Qualitative guidelines
are as follows.  Within the range of scenarios explored here, the most
important condition is the coincidence of wake pattern speed and the
disk bending-mode frequency.  Exploration suggests the 2:1 resonance
between the satellite and orbital azimuthal frequencies and the
ILR-like resonance are most important.  A secondary consideration is
the location and amplitude of the wake itself which depends on the
halo profile and the existence of low-order resonances in the vicinity
of the disk.  These two features are not independent.  However, if one
{\it could} fix the orbital frequencies of halo stars, the wake
amplitude would be proportional to the halo density and if one {\it
  could} fix the density, the wake location would scale with the
orbital frequencies.  The halos considered here are chosen to have
flat rotation curves between the outer disk and satellite pericenter.
Because the wake is dominated by the 2:1 resonance, the wake peaks at
roughly half the pericenter distance.  Therefore, increasing the mass
of the halo will tend to increase the wake amplitude but can decrease
the disk warp if the frequency match with the global disk modes is
less favorable.

Because of this complicated interplay and sensitivity to the actual
disk and halo profiles, I will illustrate the range of possibilities
with some examples rather than give an exhaustive set of models.

\subsection{Disk model dependence}

\begin{figure}
  \mbox{\epsfxsize=\colsize\epsfbox{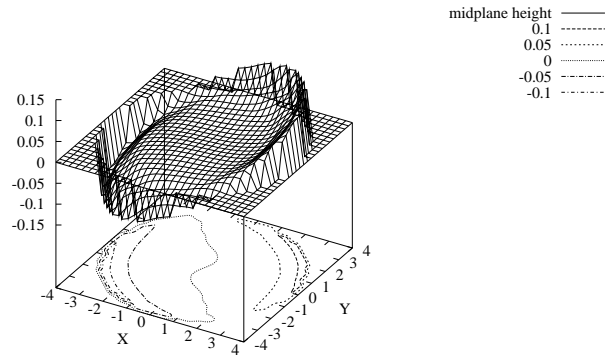}}
  \caption{Warp height for the Hunter $N=16$ disk with outer radius
    $R_{disk}=4$ and halo model $W_0=3, R=28, M=10$ (cf. Fig.
    \protect{\ref{fig:h432}}).}
\label{fig:h431}
\end{figure}

For a standard halo and satellite interaction, the disk warp in a
Hunter $N=16$ and Hunter-Toomre 16X provide a telling comparison.  The
Hunter disk has a shallower less centrally concentrated profile than
the Hunter-Toomre disk, naively suggesting that the Hunter disk will
be more susceptible to an outer disturbance.  But, because the
lower-frequency mode in the Hunter-Toomre disk better couples to
forcing by the halo wake, its response is larger.  Figure
\ref{fig:h431} shows the response for the Hunter disk for comparison
with Figure \ref{fig:h432}.  Figures \ref{fig:h437} and \ref{fig:h439}
compare the same scenario for a more centrally concentrated halo model
(King $W_0=7$, cf. \S\ref{sec:haloprof}).  As above, the Hunter $N=16$
is less warped than the Hunter-Toomre 16X disk.

\begin{figure}
  \mbox{\epsfxsize=\colsize\epsfbox{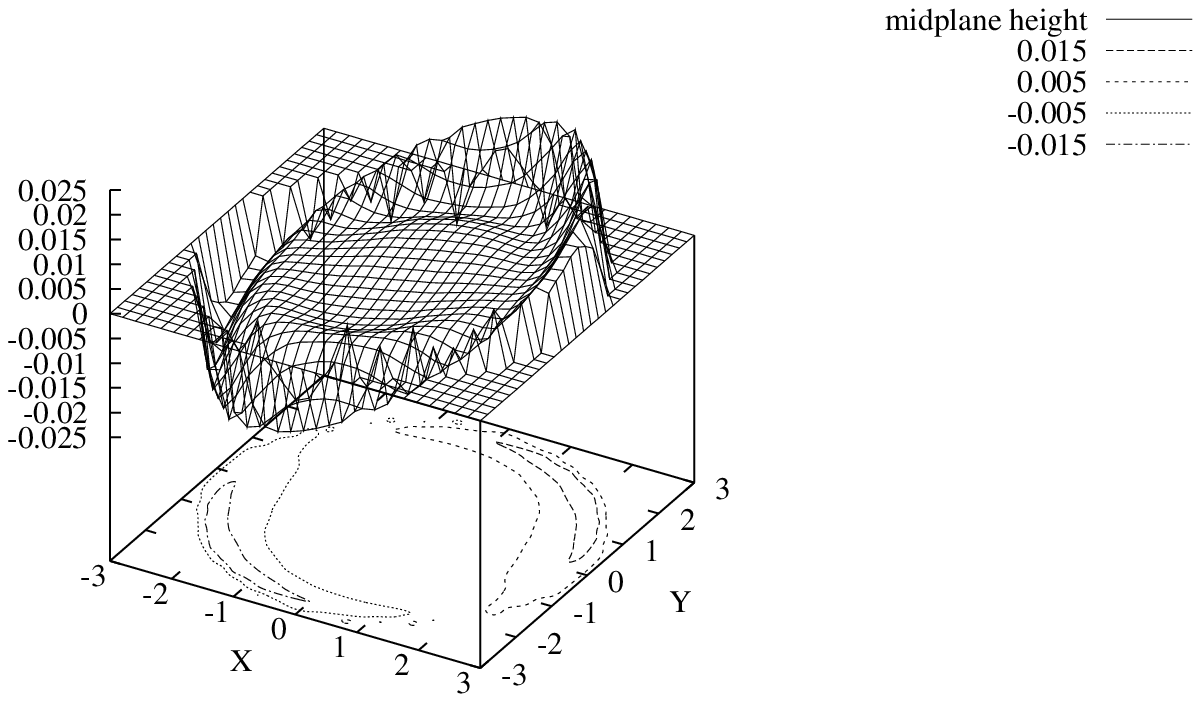}}
  \caption{Warp height for the Hunter $N=16$ disk with outer radius
    $R_{disk}=3$ and halo model $W_0=3, R=28, M=10$ (cf. Fig.
    \protect{\ref{fig:h432}}).}
  \label{fig:h442}
\end{figure}

\begin{figure}
  \mbox{\epsfxsize=\colsize\epsfbox{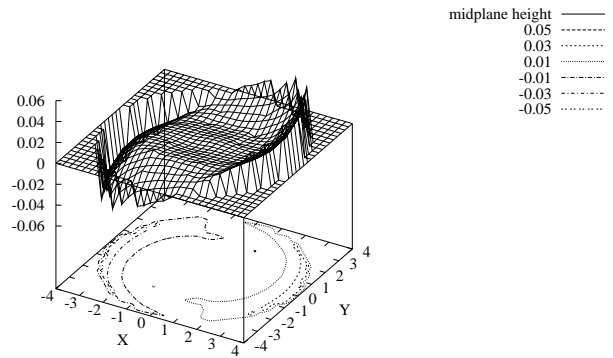}}
  \caption{Warp height for the Hunter $N=16$ disk with outer radius
    $R_{disk}=4$ and halo model $W_0=7, R=28, M=10$.}
  \label{fig:h437}
\end{figure}

\begin{figure}
  \mbox{\epsfxsize=\colsize\epsfbox{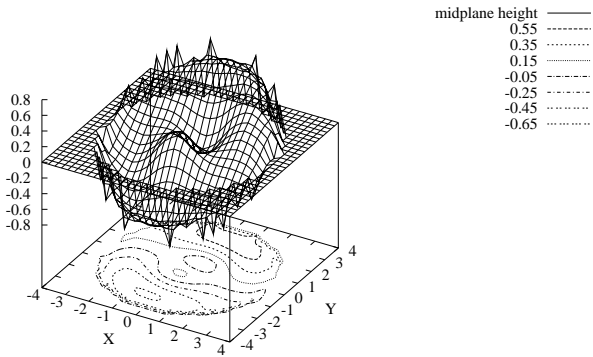}}
  \caption{Same as Fig. \protect{\ref{fig:h437}} but for the
    Hunter-Toomre 16X disk.}
  \label{fig:h439}
\end{figure}

\begin{figure}
\mbox{\epsfxsize=\colsize\epsfbox{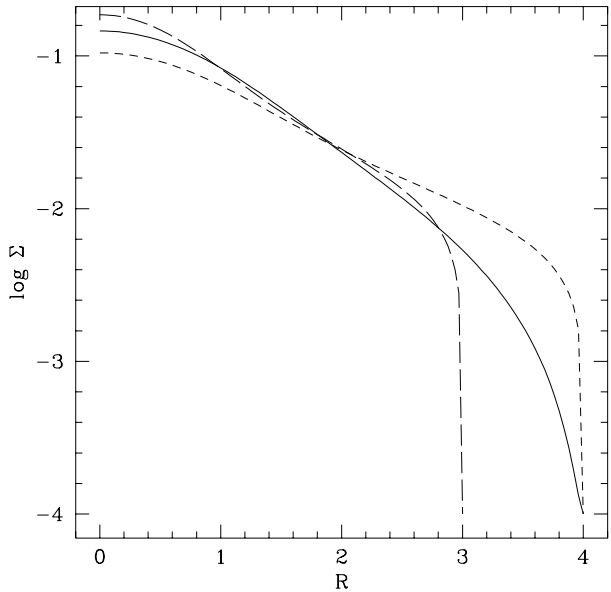}}
\caption{Comparison of the 16X disk (solid) and the Hunter $N=16$
  disks for $R_{disk}=3, 4$ (long dash, short dash).}
\label{fig:16comp}
\end{figure}

Figure \ref{fig:16comp} contrasts the the Hunter-Toomre 16X and the
Hunter $N=16$ profiles.  The latter is scaled to $R_{disk}=3$ and $4$.
Note that the Hunter $N=16$ for $R_{disk}=3$ is quite similar to the
Hunter-Toomre 16X with $R_{disk}=4$ inside of $R\approx2.8$.  Figure
\ref{fig:h442} shows that the warp in a Hunter $N=16$ disk with the
same standard halo satellite perturbation is much smaller!  Although
the disk less extended, the mode location and morphology are to blame.
The overall weaker halo support means that the tipping mode is closer
to a bodily tip and has very low frequency.  Therefore, it poorly
couples to the wake.  The first retrograde and prograde modes, which
are warp-like, have relatively high frequencies which also couple
poorly.

\subsection{Effect of halo mass and satellite orbit} \label{sec:halomass}

\begin{figure*}
  \mbox{ \mbox{\epsfxsize=\colsize\epsfbox{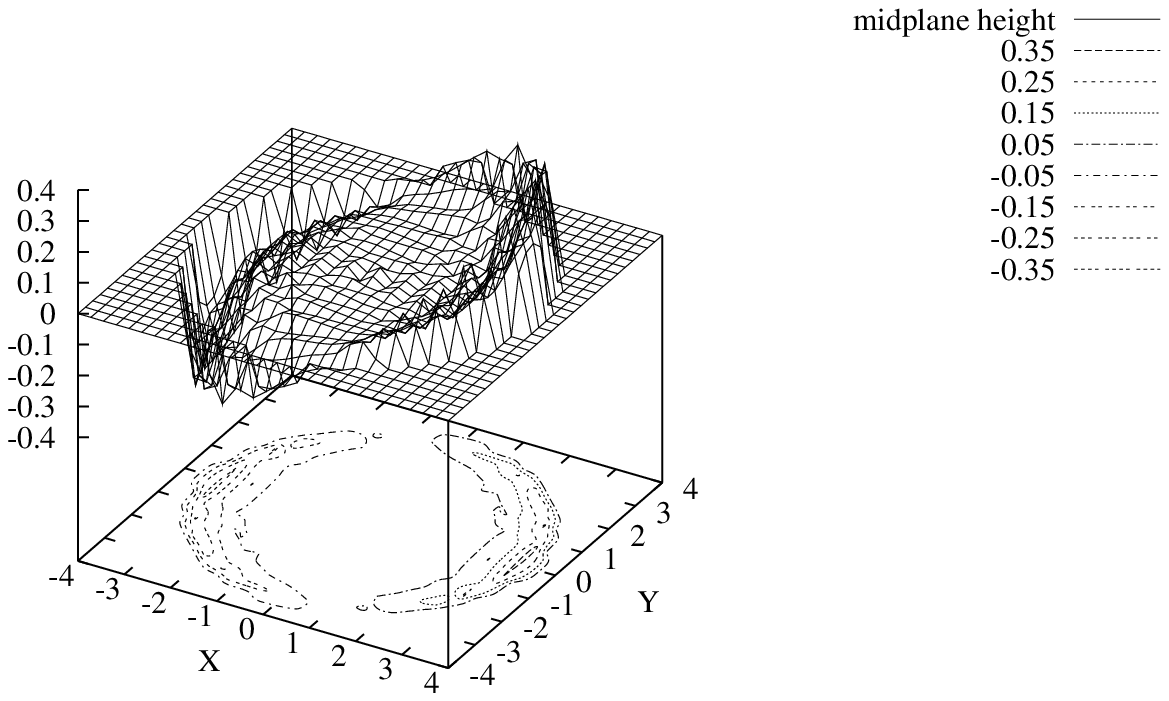}}
    \mbox{\epsfxsize=\colsize\epsfbox{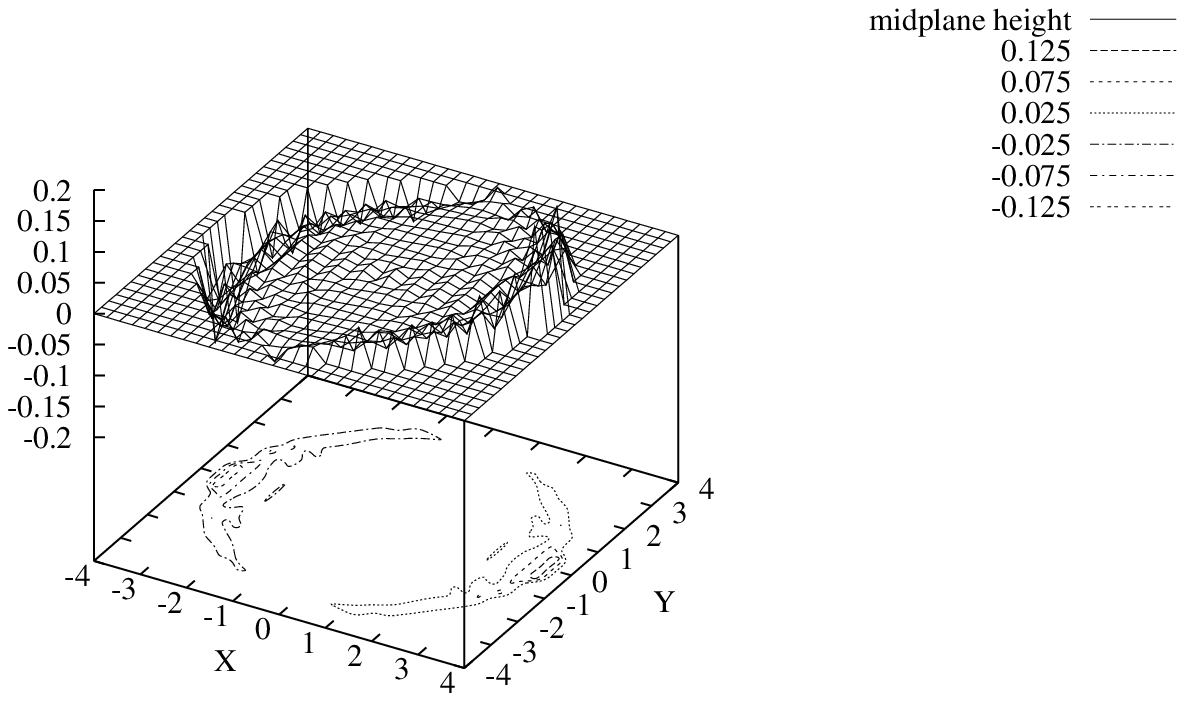}} }
\caption{Warp height for 16X disk with outer radius $R_{disk}=4$ and halo
  model $W_0=3, R=28, M=15$.  Scaled to Milky Way, 0.1 units
  corresponds to 175 pc for the large LMC mass estimate
  (cf. Fig. \protect{\ref{fig:h432}} for $M=10$).}
\label{fig:h440}
\caption{Same as Fig. Fig. \protect{\ref{fig:h440}} but for $M=20$.}
\label{fig:h441}
\end{figure*}

Figures \ref{fig:h432}, \ref{fig:h440} and \ref{fig:h441} show the
warp in the Hunter-Toomre 16X disk for increasing halo mass, $M=10,
15$ and $20$ respectively, for the standard model.  As the halo mass
increases, the amplitude of warp amplitude decreases due to a more
poorly matched halo-wake pattern speed.  In a related scenario, the
amplitude of the wake increases with decreasing satellite energy for a
fixed halo and disk as expected.  However, the warp shape changes with
energy because the increasing orbital frequencies couple differently
to the continuum disk response.

\subsection{Dependence on halo profile} \label{sec:haloprof}

Here, we explore a more concentrated halo, a $W_0=7$ King model, and
therefore a sub-maximal disk model.  The $W_0=7$ halo profile is too
centrally concentrated to be an acceptable match to standard rotation
curves: the rotation curve rises too steeply and over-supports the disk
in the inner galaxy.  The response for both the 16X and $N=16$ modes
are shown in Figures \ref{fig:h437} and \ref{fig:h439} for both disks.
The warp has a feature inside of of $R=1$ (inside the solar circle in
the Milky Way) due to the larger effect of the halo at smaller radii.
The outer warp profile is more gradual and relatively larger at
smaller radii.  Similar to the standard model, the warp in
Hunter-Toomre 16X disk is stronger than in the Hunter $N=16$ disk.

\section{N-body simulations} \label{sec:n-body}
  
The semi-analytic method of \S\ref{sec:method} is ideally suited to
N-body simulation. The same biorthogonal bases can be used to
represent the potential and force fields of a particle distribution
(e.g. Clutton-Brock 1972, 1973\nocite{Clut:72,Clut:73}, Hernquist \&
Ostriker 1992\nocite{HeOs:92}).  Because the Poisson equation is
linear, each component can be represented by an expansion suited to
its geometry.  Similarly, one can easily tailor the inter-component
forces to isolate any particular interaction.  In particular, the
n-body tests are not hampered by the difficultly in producing a
perfect time-dependent equilibrium disk-halo model; the disk feels the
background halo but halo does not react to the background disk, as
described in \S\ref{sec:combined}.

There are two differences between the simulation and the perturbation
calculation.  First, the disk must be thickened to keep it stable
against local instability (Toomre 1964\nocite{Toom:64}).  We use a
simple isothermal disk with $\sigma_z=20\kms$.  This disk would also
be bar unstable without a bulge component.  A rigid Hernquist
(1990\nocite{Hern:90}) model with scale length of 1.4 kpc and mass of
0.2 $M_{disk}$ (roughly $1.2\times10^{10}\msun$) suppressed bar
growth.  Second, the direct acceleration by the satellite
differentially accelerates the disk causing the expansion center to
drift from the center of mass.  This technical difficultly was
circumvented for the purpose of these tests by dividing the original
satellite in two and placing half of of the mass an mirror in the same
orbit but $180^\circ$ out of phase.
  
Simulations with 100,000 particles with several different partitions
between the disk and halo components were performed on a network
parallel n-body code using the LAM implementation (Burns, Daoud \&
Vaigl 1994\nocite{LAM:94}) of MPI (e.g. Gropp, Lusk \& Skjelum
1994\nocite{MPI:94}).  This particular force evaluation scheme lends
itself to workstation clusters because the communication overhead is
very low (only coefficients need to be passed, e.g. Hernquist,
Sigurdsson \& Bryan 1995\nocite{HeSB:95}) as long as the load per node
remains balanced.  To suppress transients, the satellite is slowly
turned on over an orbital period.  In general, the predicted features
were observed: vertical wakes with midplane heights of roughly 500 pc
in the outer disk in the predicted orientation.
  
Unfortunately even with 100,000 particles and the approximations above
designed to clearly isolate the $m=1$ wake, the simulation results are
difficult to interpret due to discreteness noise.  Simulations without
any satellite reveal that the noise has two effects.  First, the disk
origin random walks in the fluctuating $l=m=0$ halo component.  This
causes $m=0$ vertical distortions which lead to scale height
thickening.  One should expect these low-order fluctuations are
amplified above the Poisson amplitude by the global gravitational
response as described in Weinberg (1993,
1997\nocite{Wein:93,Wein:97}).  Secondly, one finds $m=1$ height
distortion from the $l=2$ noise-excited halo component similar in
scale to that expected from the satellite excitation.  In retrospect,
this might have been expected.  Any response to a perturbation will
feature any discrete modes.  For the halo, the strongest of these will
be weakly damped.  The halo wake, which is now the self-gravitating
response to the disk distortion, will be contain the same weakly
damped mode found in the response to the satellite.  Therefore,
particle fluctuation noise can excite a similar with modes and a
similar disk response.  The amplitude of the noise excited distortions
was as much as 30\% of the amplitude observed with a satellite
present.  However, different Monte Carlo realizations of the same
initial model gave vertical warps of arbitrary orientation whereas in
the presence of the satellite, the predicted orientation is obtained.
  
A proper n-body investigation of these effects will require
simulations with many more particles and will be left for the future.
This investigation, however, inadvertently raises the interesting
possibility that a satellite may interact with noise-excited features
to produce an overall response which is larger than the
time-asymptotic predictions described above.  Because the modes are
already excited by noise, they may be entrained and strengthed by the
external perturbation.  This is similar to other noise amplification
mechanisms found in Nature, e.g. {\it stochastic resonance} mechanism
observed in neurobiology (e.g.  Collins, Chow \& Imhoff
1995\nocite{CoCI:95}).  Moreover, there are many sources of
inhomogeneity in real galaxies such as star clusters, gas clouds,
unmixed streams from dwarf dissolution, and other sources of
perturbations such as tidal distortions from interactions with a host
galaxy cluster and near-neighbor interactions.  In fact, the noise in
the simulation presented here corresponds to that produced by a halo
of $10^6M_\odot$ black holes (e.g. Lacey \& Ostriker
1985\nocite{LaOs:85}).  Altogether, it seems likely that observable
effects on disks due to halo interactions will be stronger than the
predictions in \S\ref{sec:results}.

\section{Summary and future work} \label{sec:summary}

This paper summarizes the effect of a satellite companion on a
galaxian disk embedded in a responsive (or {\it live}) halo.  Because
the mechanisms described here involve multiple time and length scales,
the interaction is complex and depends on the details of each
dynamical system: satellite orbit, halo and disk structure.  The
general conclusions and expectations are enumerated below:

\begin{enumerate}
  
\item The halo can sustain a significant wake in the presence of an
  LMC-like satellite.  This wake is most often caused by a 2:1 orbital
  resonance and therefore peaks roughly half way between the satellite
  orbit in the halo center.
  
\item The halo wake, because it has structure at smaller
  galactocentric radii, can excite warp modes in the disk more
  efficiently than direct tidal perturbation.
  
\item The strongest warps obtain for a near commensurability between a
  disk mode and the pattern speed of the halo wake. This makes robust
  predictions for warp amplitudes more difficult, but we find that
  warps with observable amplitude can easily result by the mechanism
  described here.
  
\item For halos with roughly flat rotation curves out to roughly 50
  kpc, the satellite orbit (pericenter, plane orientation, and
  eccentricity) which determines the forcing frequencies and disk
  profile which determines the bending frequencies are most important
  in determining the induced warp.
  
\item A polar satellite orbit will produce the largest warp.  The
  inferred LMC orbit is nearly optimal for maximum warp production.
  
\end{enumerate}
The halo wake (e.g.  Figs. \ref{fig:halowake1} and
\ref{fig:halowake2}) plays a critical role in producing a warp and is
an observable consequence of the massive dark halo.  A warp survey
combined with recent satellite surveys (Zaritsky et al.
1993\nocite{ZRFW:93}) may provide additional statistical evidence for
the massive halo hypothesis.

Several possibly important interactions have been ignored in this
study and present topics for further research.  First, the disk is
model is infinitely thin; including the three-dimensional structure
will damp the warp producing modes but this damping is likely to be
small at large scales (Weinberg 1991\nocite{Wein:91b}).  Second, the
disk respond to the halo wake and the halo responds to the
two-dimensional disk distortion but not the three-dimensional one.
Therefore the dynamical friction against the halo explored by Nelson
\& Tremaine (1995\nocite{NeTr:95}) is not included.  The calculational
method used here is tractable because of the assumption that all
transients have mixed away.  As they mix, they produce fluctuations on
many scales.  Intrinsic halo inhomogeneities such as clouds, clusters
and massive black holes are also a source of noise.  Together, this
noise may seed interesting observable features, such as inner bars and
arms, which are not part of the long-term wake.  In addition, as
described in \S\ref{sec:n-body}, these fluctuations will drive the
same modes which produce the wakes at largest scales.  N-body
simulations suggest that noise-excited structure can have a effect on
the disk similar to the satellite companion.  This leads to the
possibility that the amplitude of the large-scale response to an
external disturbance may be amplified by entraining the pre-existing
noise-excited features.

\section*{Acknowledgements}
I thank Peter Goldreich, Mark Heyer, Neal Katz, Ron Snell, and Scott
Tremaine for helpful discussion.  This work was supported in part by
NSF AST-9529328 and the Sloan Foundation.

\appendix

\section{Computing the disk distribution function} \label{sec:diskDF}

The distribution function is the solution to the following integral
equation:
\begin{equation} \label{eq:dfint}
\Sigma(R) = \int d^2v f(E, J),
\end{equation}
where $E$ and $J$ are the orbital energy and angular momentum,
respectively.  For a given energy, the maximum angular momentum, that
of a circular orbit with energy $E$, is denoted by $J_{max}(E)$.  The
potential of the halo ($\Phi_{halo}$) and disk profile ($\Sigma(R)$,
$\Phi_{disk}$) are fixed and assumed to be known to start.  We allow
the distribution function to be represented by a Gaussian basis in the
variables $E$ and $\kappa\equiv J/J_{max}(E)$:
\begin{eqnarray}
{\tilde f}(E, \kappa) &=& \sum_{i,j} a_{ij} \exp\left[
        -(E-E_i)^2/2\sigma_E^2 -(\kappa-\kappa_j)^2/2\sigma_\kappa^2
        \right]. \nonumber \\
\end{eqnarray}
At any point $R_k$, the distribution function is related to
$\Sigma(R_k)$ through equation (\ref{eq:dfint})
\begin{equation}
{\tilde\Sigma}(R_k) =   2\int_0^{v_{max}(R_k)} dv_r
                 \int_0^{\sqrt{v_{max}^2(R_k) - v_r^2}} dv_t
                {\tilde f}(E, \kappa)
\end{equation}
where $v_{max}(R_k) = \sqrt{2(E_{max}-\Phi_{disk}-\Phi_{halo})}$,
$E=(v_r^2 + v_t^2)/2 + \Phi_{disk}+\Phi_{halo}$, and
$\kappa=R_kv_t/J_{max}(E)$.  An exact solution requires that the
$a_{ij}$ to satisfy equation (\ref{eq:dfint}) for all $R$.  We can
state this demand as the set $a_{ij}$ which minimizes the square of
the difference of $\Sigma(R_k)$ and ${\tilde\Sigma}(R_k)$ at all $R$.
This demand may be discretized to the set $a_{ij}$ which minimizes
\begin{eqnarray} \label{eq:chi2}
  \chi^2 &=& \sum_k w_k \left[\Sigma(R_k) -
    {\tilde\Sigma}(R_k)\right]^2, \nonumber \\
  &=& \sum_k w_k
  \left[\sum_{i,j}a_{ij}\Sigma_{ij}(R_k) -
    {\tilde\Sigma}(R_k)\right]^2 
\end{eqnarray}
with the condition
\begin{equation}        
        {\tilde f}(E, \kappa) \ge 0
\end{equation}
where $w_k$ is a weighting, which may be $w_k=1$.  This defines a
standard quadratic programming problem for the $a_{ij}$ which we solve
using Powell's algorithm (1982) as implemented in the QLD code
provided by Andre Tits (Lawrence, Zhou \& Tits 1994\nocite{CFSQP:94}).

Distribution functions used for models described here used a grid of
$20\times 20$ in $E$ and $\kappa$ and penalized the expression in
equation (\ref{eq:chi2}) to construct a tangentially anisotropic
distribution,
\begin{eqnarray} \label{eq:chi2P}
  \chi^2_p &=& \sum_k w_k \left\{
    \left[\sum_{i,j}a_{ij}\Sigma_{ij}(R_k) -
      {\tilde\Sigma}(R_k)\right]^2 
    + \right.
    \nonumber \\ && \left.
    \lambda \sum_{i,j} \sum_{r,s} a_{ij}a_{rs} 
    {\kappa_j\kappa_s}^{-\alpha}\right\},
\end{eqnarray}
with $w_k=1$, $\lambda=10^{-3}$ and $\alpha=6$.

\section{Perturbation coefficients for an orbiting satellite}
\label{sec:pertcoef}

The biorthogonal expansion coefficients (cf. eq. \ref{eq:satexp}) for
a satellite orbiting in a spherical halo are conveniently derived by
expanding its gravitational potential in an action-angle series.
Assuming a point-mass perturber, the coefficients are
\begin{equation} 
  b_i^{lm}(t) = Y^\ast_{lm}(\theta(t), \phi(t)) p^{lm\,\ast}_i(r(t)).
  \label{eq:ptexpandA}
\end{equation}
Because ${\bf r}(t)$ is quasiperiodic in two frequencies for a
spherical halo, we may expand equation (\ref{eq:ptexpandA}) (or eq.
\ref{eq:ptexpand} in the main text) in a Fourier series in time.

This is straightforwardly done in the perturber's orbital plane and
then rotated to the desired orientation using the rotational
properties of spherical harmonics (e.g. Edmonds 1960\nocite{Edmo:60}):
\begin{equation} \label{eq:erot}
  Y_{lk}(\pi/2,\psi)  = \sum_m D^{l\,\ast}_{km}(\alpha,\beta,\gamma)
  Y_{lm}(\theta, \phi)
\end{equation}
where
\begin{equation}
  D^l_{km}(\alpha,\beta,\gamma) = e^{ik\alpha}r^l_{km}(\beta)e^{im\gamma}.
\end{equation}
The $r^l_{mk}(\beta)$ are the rotation matrices and $\alpha$, $\beta$,
$\gamma$ are the Euler angles describing the orientation to the
orbital plane.

Using this and the inverse of equation (\ref{eq:erot}), we can now
expand equation (\ref{eq:ptexpandA}) in action-angle variables. For
spherical systems, the third angle variable describes the line of
nodes; it has zero frequency and has been suppressed.  The
action-angle coefficients are then defined by
\begin{eqnarray}
  b^{lm}_{i{\bf l}} &=& {1\over(2\pi)^2}\sum_k \int d^3w\, e^{-i(l_1w_1 +
    l_2w_2)} e^{im\gamma} r^l_{mk}(\beta) \times \nonumber \\ &&
  e^{ik\alpha} Y_{lk}(\pi/2,0)e^{ik\psi} p^{lm\,\ast}_i(r(w_1)) \nonumber \\
  &=& {1\over2\pi}\sum_k Y_{lk}(\pi/2,0) r^l_{mk}(\beta)
  \delta_{kl_2}\delta_{ml_3} e^{im\gamma} r^l_{mk}(\beta) e^{ik\alpha}
  \times \nonumber\\
  && \int dw_1 e^{-i(l_1w_1 - l_2(\psi-w_2))} p^{lm\,\ast}_i(r(w_1)). 
\end{eqnarray}
The angles $w_1$ and $w_2$ describe the equal-time motion from
pericenter to pericenter and the mean azimuthal motion respectively.
The angle $\alpha$ describes the rotation of the orbital plane;
$\alpha=0$ corresponds to the $y$ axis in the orbital plane coincident
with the line of nodes.  The angle $\gamma$ describes the orientation
of the line of nodes in the original azimuthal coordinate; $\gamma=0$
places the line of nodes coincident with the $y$ axis in the original
system.  Because the orbital radius and the deviation of the true
azimuth from the mean azimuth ($\psi-w_2$) is even in $w_1$, the
integral over $w_1$ may be simplified and one finds:
\begin{equation}
  b^{lm}_{i{\bf l}} =  e^{im\gamma} e^{il_2\alpha}
  Y_{ll_2}(\pi/2,0) r^l_{l_2m}(\beta)
  W^{l_1\,i\,\ast}_{l\,l_2\,m}({\bf I}),
\end{equation}
where $W$ is defined by
\begin{equation}
  W^{l_1\,i}_{l\,l_2\,m}({\bf I}) = {1\over\pi} \int_0^\pi dw_1\,
  \cos(l_1w_1 +l_2f(w_1))p^{lm}_i(r(w_1))
\end{equation}
with
\begin{equation}
f(w_1) = \oint dr \left[2(E-\Phi_0)-J^2/r^2)\right]^{-1/2}
\left(\Omega_2-J/r^2\right).
\end{equation}
Each term in the expansion in equation (\ref{eq:satexp}) is then
\begin{equation} 
  b_i^{lm}(t) = \sum_{l_1, l_2=-\infty}^\infty e^{im\gamma}
  e^{il_2\alpha}
  Y_{lm}(\pi/2, 0) W^{l_1\,i\,\ast}_{l\,l_2\,m}
  e^{i(l_1\Omega_1+l_2\Omega_2)t}
\end{equation}
for $w_{10}=w_{20}=0$.  We will focus on the $m=1,2$ responses for
$l\le4$.  Several tests with $l\le6$ suggest that $l\le4$ dominate the
large-scale response.

\section{Numerical computation of response matrices} \label{sec:numresp}

The elements of the response matrix have been derived in several
places (Paper II and references) and are given by:
\begin{eqnarray}
        {\cal R}^{lm}_{ij}(s) &=& -4\pi (2\pi)^3 {2\over 2l+1}
        \sum_j\sum_{\bf l}
        \int {dE\,dJ\,J\over\Omega_1(E,J)} i{\bf l}\cdot{\partial
        f_o\over\partial{\bf I}}
        \times  \nonumber \\
        && {1\over s+i{\bf l}\cdot{\bf\Omega}} |Y_{ll_2}(\pi/2,0)|^2 
        W_{ll_2m}^{l_1i\ast}({\bf I}) W_{ll_2m}^{l_1j}({\bf I})
        .
        \label{eq:respmat}
\end{eqnarray}
The inverse Laplace transform as described in \S\ref{sec:method}
requires evaluation of matrix elements of this form in two contexts.
The Laplace transform of the coefficients in equation
(\ref{eq:satfreq}), ${\hat b}^{lm}_i(s)$, contribute a simple pole on
the real axis is easily performed by deforming the contour to $\Im
(s)\rightarrow-\infty$.  The inverse transform then has the following
form
\begin{eqnarray}
  {\bf a}^{lm}(t) &=& {1\over2\pi i}\int^{c+i\infty}_{c-i\infty}
  ds {\bf{\cal D}}^{-1}(s){\cal R}(s)\cdot{\bf{\hat b}}^{lm}
  \nonumber\\
  &=& {1\over2\pi i}\int^{c+i\infty}_{c-i\infty}
  ds e^{st} {\bf{\cal D}}^{-1}(s){\cal R}(s)\cdot{\bf b}^{lm} {1\over
    s-i\omega}.
  \label{eq:deform}
\end{eqnarray}
Although the stability assumption ensures that ${\bf\cal D}^{-1}$ has
no poles in the $\Re(s)>0$ half plane, there are poles in the
$\Re(s)<0$ half plane corresponding to damped modes (e.g.  Weinberg
1994\nocite{Wein:94}).  Of course, weakly damped modes will be
subdominant to an oscillatory mode after sufficiently long time and we
assume this limit here.

The elements of ${\bf{\cal D}}(s)$ for the deformed integration path
must be analytically continued to $\Re(s)\le 0$ from $\Re(s)>0$ where
the transform is defined.  This leads to a Cauchy integral with two
simple poles on the real axis for each matrix element (cf. eq.
\ref{eq:respmat}): $s=i\omega$ and $s=i{\bf l}\cdot{\bf\Omega}$.
Numerically, this may be straightforwardly evaluated by subtracting
the singularity from the integrand and evaluating it separately.
Consider the following Cauchy integral:
\begin{equation}
  \int dz {f(z)\over z-z_o} = \int dz {f(z) - f(z_o)\over z-z_o} +
  f(z_o) \int dz {1\over z-z_o}.
\end{equation}
The first term on the right hand side is non-singular and can be
evaluated by simple quadrature.  The second term can be performed
analytically once the contour is specified.  In this case, analytic
continuation requires the standard Landau contour (cf. Krall \&
Trivelpiece 1973\nocite{KrTr:73}).  Unlike the case of damped modes,
the evaluation of the forced response requires no extrapolation or
explicit complex integration.

\bsp
\label{lastpage}
\end{document}